\documentclass[lettersize,journal]{IEEEtran}
\usepackage{amsmath,amsfonts}
\usepackage{algorithm}
\usepackage{algpseudocode}
\usepackage{amssymb}
\usepackage{amsfonts}
\usepackage{array}
\usepackage[caption=false,font=footnotesize,labelfont=rm,textfont=rm]{subfig}
\usepackage{textcomp}
 \usepackage{url}
\usepackage{bm}
\usepackage{subeqnarray}
\usepackage{fancyhdr}
\usepackage{textgreek}
\usepackage{xcolor}
 \usepackage{url}
\usepackage{verbatim}
\usepackage{graphicx}
\usepackage{cite}
\usepackage{adjustbox}
\usepackage{stfloats}
\usepackage{subfloat}
\usepackage{pifont}
\usepackage{multirow}
\usepackage{booktabs}
\usepackage{threeparttable}
\begin{document}
\title{Joint Optimization based on Two-phase GNN in RIS- and DF-assisted MISO Systems with Fine-grained Rate Demands}

\author{Huijun Tang,~\IEEEmembership{Member,~IEEE}, Jieling Zhang, Zhidong Zhao, Huaming Wu,~\IEEEmembership{Senior Member,~IEEE}, Hongjian Sun,~\IEEEmembership{Senior Member,~IEEE} and Pengfei Jiao,~\IEEEmembership{Member, IEEE}
\IEEEcompsocitemizethanks{\IEEEcompsocthanksitem H. Tang, J. Zhang, Z. Zhao and P. Jiao are with the School of Cyberspace, Hangzhou Dianzi University, Hangzhou 310018, China. Email: \{tanghuijune,232270031, zhaozd, pjiao\}@hdu.edu.cn.\protect}
\IEEEcompsocitemizethanks{\IEEEcompsocthanksitem H. Wu is with the Center for Applied Mathematics, Tianjin University, Tianjin 300072, China. E-mail: whming@tju.edu.cn.\protect}
\IEEEcompsocitemizethanks{\IEEEcompsocthanksitem H. Sun is with the Department of Engineering, Durham University, Durham DH1 3LE, UK. E-mail: hongjian.sun@durham.ac.uk.\protect
}
\thanks{
(Corresponding author: Huaming Wu)}
\thanks{For the purpose of open access, the author has applied a Creative Commons Attribution (CC
BY) license to any Author Accepted Manuscript version arising.}	}

\markboth{IEEE Transactions on Wireless Communications,~Vol.~, No.~, ~2024}%
{Shell \MakeLowercase{\textit{et al.}}: A Sample Article Using IEEEtran.cls for IEEE Journals}

\maketitle

\begin{abstract}
Reconfigurable intelligent Surfaces (RIS) and half-duplex decoded and forwarded (DF) relays can collaborate to optimize wireless signal propagation in communication systems. Users typically have different rate demands and are clustered into groups in practice based on their requirements, where the former results in the trade-off between maximizing the rate and satisfying fine-grained rate demands, while the latter causes a trade-off between inter-group competition and intra-group cooperation when maximizing the sum rate. However, traditional approaches often overlook the joint optimization encompassing both of these trade-offs, disregarding potential optimal solutions and leaving some users even consistently at low date rates. To address this issue, we propose a novel joint optimization model for a RIS- and DF-assisted multiple-input single-output (MISO) system where a base station (BS) is with multiple antennas transmits data by multiple RISs and DF relays to serve grouped users with fine-grained rate demands. We design a new loss function to not only optimize the sum rate of all groups but also adjust the satisfaction ratio of fine-grained rate demands by modifying the penalty parameter. We further propose a two-phase graph neural network (GNN) based approach that inputs channel state information (CSI) to simultaneously and autonomously learn efficient phase shifts, beamforming, and relay selection. The experimental results demonstrate that the proposed method significantly improves system performance.
\end{abstract}


\section{Introduction}
\IEEEPARstart{T}he exponential growth of wireless data traffic, which is driven by the proliferation of smart devices and the advent of the Internet of Things (IoT), has posed substantial challenges to existing communication infrastructures~\cite{4,10195222,10210075,8954683}. Traditional methods to address these challenges, such as deploying additional base stations and leveraging more spectrum, have often entailed high costs and increased energy consumption.~\cite{1,2,10040976}. Reconfigurable intelligent Surfaces (RIS), which intelligently reconfigures the propagation environment through passive, low-power, and controllable surfaces, has gained significant attention as a compelling alternative in wireless communication research. 

\begin{figure}[ht]
\centering
\adjustbox{width=8.6cm,height=4.6cm,keepaspectratio}{\includegraphics{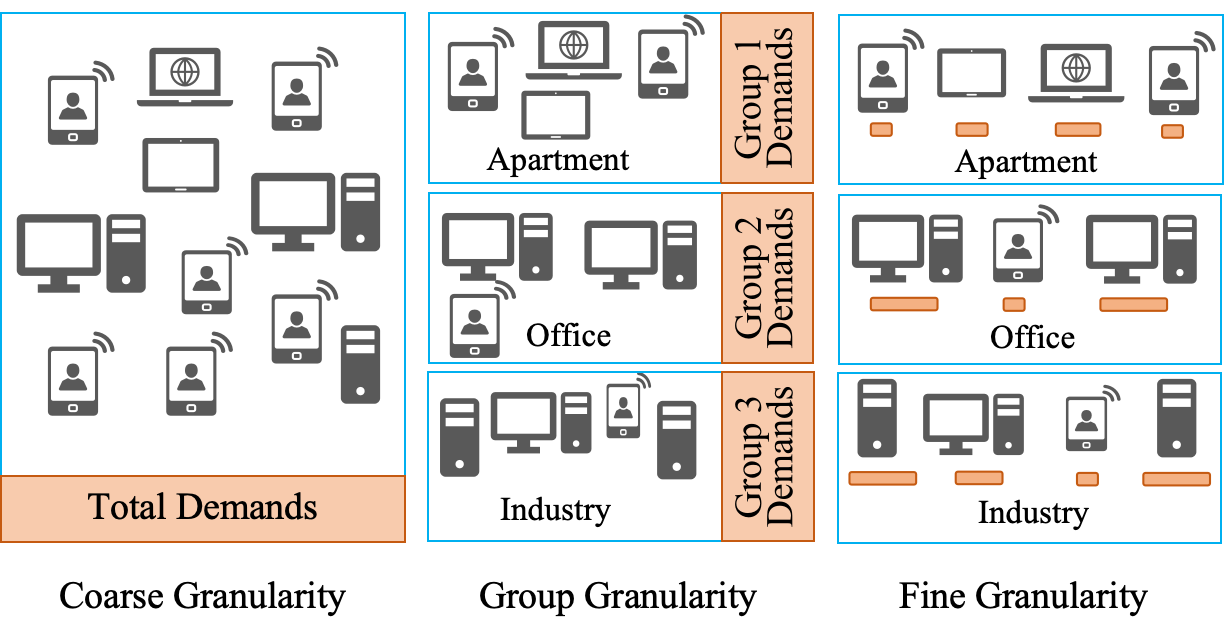}}
\vspace{-0.1in}
\caption{Classify device rate demands into three granularities: coarse-grained, group-grained, and fine-grained.}
\label{fine}
\vspace{-0.1in}
\end{figure}

An RIS comprises a set of low-cost and reflecting elements capable of manipulating electromagnetic waves incident upon them, thereby controlling, amplifying, or attenuating electromagnetic waves. These reflective components, controllable and configurable, provide superior communication objectives compared to conventional wireless systems~\cite{3,9950621,9780612,9337218,10836876}. Simultaneously, the relay decodes, remodulates, and retransmits signals, thereby mitigating error propagation and enhancing signal reliability. Additionally, deploying a relay also for flexible network expansion to maintain stable connectivity across all areas~\cite{8490219,10072922,9440847}. Consequently, in 6G communications, the hybridization of RIS and relays has swiftly garnered attention, with the objective of enhancing wireless network coverage and signal integrity through their integration. By dynamically adjusting the reflection coefficients of the RIS and the amplification factor of the relay, a more flexible and efficient communication link has been realized in~\cite{6758216}. The researchers conducted an in-depth analysis of the performance of the relay-assisted reflection intelligent surface network. The research focused on evaluating the effectiveness of RIS under different relay strategies, including amplified forwarded (AF) and decoded and forwarded (DF) relays. The results show that a well-configured RIS enhances signal quality and mitigates system complexity. This technology leverages the synergies of RIS and relays, offering extended coverage and active relay signal processing, along with the energy efficiency and cost benefits inherent to RIS~\cite{17,18}. 

Optimizing both the RISs and relays can enhance the coverage and data transmission rate of the communication system by leveraging the signal amplification characteristics of RISs' beamforming and relaying to overcome signal attenuation and transmission distance limitations. This approach is applicable to various communication scenarios, such as indoor/outdoor environments and mobile/fixed communication setups. Tran \textit{et al.}~\cite{9807389} utilized multi-RISs with a relay to enhance the performance of a low-power wide-area network. Nguyen \textit{et al.}~\cite{9733238} explored how reflective smart surfaces can be integrated with relays to adapt to complex wireless environments. Therefore, research in joint optimization for both the RISs and relays is of paramount importance as it significantly enhances communication system performance and user experience~\cite{9426939,9756313}. To unleash the potential of RIS and relay-assisted multi-antenna systems in supporting high data rates, beamforming has been optimized through various works. For instance, Guo \textit{et al.}~\cite{5} proposed an algorithm grounded in block coordinate descent (BCD) that aims to simultaneously optimize both transmit beamforming and RIS phase shifts, thereby achieving the highest possible sum rate for RIS-assisted communication systems. Hu \textit{et al.}~\cite{6} addressed the challenge of maximizing the weighted sum rate in non-convex scenarios by employing a combination of fractional programming (FP) and alternating optimization (AO) techniques.

\begin{table*}[ht!]
  \centering
  \caption{The qualitative comparison of the current literature. The symbol '\checkmark' means that the factor is taken into account, and the symbol '\ding{55}' means not taking this factor into account}
  \begin{threeparttable}   
    \begin{tabular}{|c|p{1.3cm}<{\centering}|p{1.3cm}<{\centering}|p{1.3cm}<{\centering}|p{1.7cm}<{\centering}|p{1.3cm}<{\centering}|p{2.5cm}<{\centering}|c|}
    \hline
    \textbf{Approaches}& \textbf{BS beamforming} & \textbf{RIS phase shifts} & \textbf{Multiple RISs} & \textbf{Relay beamforming} & \textbf{Relay selection} & \textbf{Fine-grained rate demands} & \textbf{Method} \\
    \hline
    Zaid \textit{et al.}\cite{17} & \ding{55} & \ding{55} & \checkmark & \checkmark & \checkmark & \ding{55} & SDP\tnote{1}, PSO\tnote{2}\\
    \hline
    Zaid \textit{et al.}\cite{18} & \ding{55} & \checkmark & \ding{55} & \checkmark & \ding{55} & \ding{55} & EPA\tnote{3}, OPA\tnote{4}\\
    \hline
    Ahn \textit{et al.}\cite{9621150} & \checkmark & \checkmark & \ding{55} & \ding{55} & \ding{55} & \ding{55} & DNN\tnote{5} \\
    \hline
    Xu \textit{et al.}\cite{9814839} & \ding{55} & \checkmark & \checkmark & \ding{55} & \ding{55} & \ding{55} & LSTM\tnote{6} \\
    \hline
    Yang \textit{et al.}\cite{9322615} & \checkmark & \checkmark & \ding{55} & \ding{55} & \ding{55} & \ding{55} & DRL\tnote{7} \\
    \hline
    Asmaa \textit{et al.}\cite{10060056} & \checkmark & \checkmark & \ding{55} & \ding{55} & \ding{55} & \ding{55} & DRL\tnote{7} \\
    \hline
    Xu \textit{et al.}\cite{9779399} & \checkmark & \checkmark & \ding{55} & \ding{55} & \ding{55} & \ding{55} & DRL\tnote{7} \\
    \hline
    Tao \textit{et al.}\cite{9427148} & \checkmark & \checkmark & \ding{55} & \ding{55} & \ding{55} & \ding{55} & GNN\tnote{8} \\
    \hline
    Chen \textit{et al.}\cite{10184122} & \checkmark & \checkmark & \ding{55} & \checkmark & \ding{55} & \ding{55} & GNN\tnote{8} \\
    \hline
    Wang \textit{et al.}\cite{14} & \checkmark & \checkmark & \ding{55} & \ding{55} & \ding{55} & \ding{55} & FP\tnote{9}, SDR\tnote{10}, SROCR\tnote{11}\\
    \hline
    Ma \textit{et al.}\cite{15} & \checkmark & \checkmark & \ding{55} & \ding{55} & \ding{55} & \ding{55} & FP\tnote{9}\\
    \hline
    Huang \textit{et al.}\cite{9344820} & \checkmark & \checkmark & \ding{55} & \ding{55} & \checkmark & \ding{55} & DRL\tnote{7}\\
    \hline
    \textbf{Ours} & \checkmark & \checkmark & \checkmark & \checkmark & \checkmark & \checkmark & GNN\tnote{8} \\
    \hline
    \end{tabular}%
    \footnotesize 
    \tnote{1}SDP: semidefinite programming, \tnote{2}PSO: particle swarm optimization, \tnote{3}EPA: equal power allocation, \tnote{4}OPA: optimal power allocation,
    \tnote{5}DNN: deep neural network,\tnote{6}LSTM: long short-term memory,\tnote{7}DRL: deep reinforcement learning,\tnote{8}GNN: graph neural network, \tnote{9}FP: fractional programming, \tnote{10}SDR: semidefinite relaxation, \tnote{11}SROCR: sequential rank-one constraint relaxation.             
    \end{threeparttable} 
  \label{tab:addlabel}%
  \vspace{-0.15in}
\end{table*}

However, these prior works overlook user aggregation and spatial distribution, which are crucial for the optimization of the RIS-relay integration, since beamforming techniques rely on the positions of users to adjust the phases and magnitudes of antennas to ensure that the signals precisely target the intended user locations. In practice, users and devices are often clustered in various geographic areas such as office buildings, factories, and residential apartments. As shown in Fig.~\ref{fine}, we categorize various locations into distinct groups and classify device date rate demands into three granularities: coarse-grained, group-grained, and fine-grained. Prior works have focused on coarse-grained optimization, but they often overlook the impact of geographical distribution on beamforming. The group granularity takes geographical distribution into account. In particular, if one particular group's total rate demand is high, the decision-making process will prioritize that group, which is called intra-group cooperation. Conversely, differing geographical locations among groups lead to rate demand-induced resource competition, termed inter-group competition. These groups exhibit diverse characteristics, leading to varying data rate demands among different devices. 

For instance, industrial IoT devices in factories may require highly reliable low-latency connections, whereas consumer electronics in residential areas may demand a high data throughput for their such as streaming and gaming. In the current 6G communication scenario, the spatial distribution of user devices is highly heterogeneous. For example, sensors in industrial IoT are densely distributed in factory areas, requiring low-latency and highly reliable connections; while streaming media devices in residential areas require high throughput. Traditional optimization methods treat users as uniformly distributed individuals, ignoring their grouping characteristics and geographic correlation, resulting in inefficient resource allocation. In addition, although existing deep neural network (DNN) based methods can handle high-dimensional channel state information (CSI), it is difficult to capture the topological dependencies among users, RISs, and relays, especially when multiple groups of users are competing for resources, and DNNs are not globally-aware enough and are prone to fall into a local optimum. This heterogeneity in spatial distribution and service requirements necessitates sophisticated network management strategies, while group-grained granularity neglects individual device requirements.

Considering the impact of the heterogeneity in spatial distribution and rate demands, we propose the joint optimization with fine-grained demands (JOFD) method, aiming to maximize the sum rate while simultaneously meeting the specific needs of each device, which can prevent users from monopolizing excessive resources and enable accurate and fine-grained management of network resources. The main contributions of this paper are summarized as follows:

\begin{itemize}
\item 
\emph{Inter- and Intra-group sum rate joint optimization}: This paper proposes an innovative optimization model for RIS- and DF-assisted MISO systems to achieve optimal strategy in sum rate after making a trade-off between inter-group competition and intra-group cooperation. Due to clustering, optimization outcomes within the same group tend to confusing, while the between positions fosters competition among different groups. To our knowledge, this is the first study addressing joint sum rate optimization that incorporates both inter-group rivalry and intra-group collaboration.
\item
\emph{Fine-grained rate demands gratification}: We propose a device-grained demands guarantee algorithm to satisfy the different rate demands of devices, which affords a more comprehensive exploration of the optimization space than coarse granularity. To achieve this, we design a novel loss function with a penalty term to optimize the sum rate and ensure fine-grained rate demands, where the satisfaction ratio of fine-grained rate demands can be tuned by modifying the penalty parameter.
\item \emph{Model and Algorithm Design}: We innovatively model the RIS- and DF-assisted MISO system as a graph and introduce a two-phase GNN to solve the optimal strategy for the active beamforming at the base station (BS) and DF relays, the passive beamforming at the RIS and the selection of the relays.Simulation data indicates that the proposed GNN-based joint optimization approach surpasses conventional methods, and the accuracy and robustness of the network are verified by the test data set. Furthermore, the proposed GNN exhibits strong generalization capabilities across varying numbers of users. 
\end{itemize}

The remainder of the paper are organized as follows: Section \ref{related} reviews the related work, Section \ref{system_model} details the system model and the problem formulation, Section \ref{method} describes the proposed GNN-based architecture, Section \ref{result} presents the simulation results for the proposed approach, and Section \ref{conclusion} concludes the paper.

\vspace{-0.1in}
\section{Related Work}
\label{related}
In recent years, the optimization of sum rate and energy efficiency in RIS-aided communication systems has garnered significant attention. Traditional methods primarily focus on leveraging mathematical optimization techniques to enhance system performance. For example, convex optimization and iterative algorithms are extensively studied to maximize the sum rate by optimizing the phase shifts of RIS elements~\cite{5,6}. An efficient channel estimation method is designed using the sparsity and correlation of the channel, and combined with a discrete phase offset passive beamforming strategy~\cite{9614196}, or by reducing pilot overhead, the transmission performance of the system is improved, which is suitable for high-frequency band scenarios such as millimeter wave communications~\cite{9952197}. However, the numerical algorithm of traditional methods still results in high computational complexity.

Machine learning (ML) techniques have become prominent for tackling optimization issues in RIS-assisted systems, presenting reduced complexity in contrast to model-driven methods. Supervised learning techniques, in particular, have shown promise in predicting the optimal configuration of RIS elements. Ahn et al. \cite{9621150} proposed a deep neural networks model that jointly optimizes beamforming and RIS phase shifts. This approach utilizes neural networks to determine optimal vectors and matrices, enhancing signal reception quality. Xu et al. \cite{9814839} proposed a three-phase joint channel decomposition and prediction framework based on deep learning by optimizing the phase offset and channel state information acquisition of RIS, which solves the problems of channel estimation accuracy as well as channel decomposition in RIS-assisted MU-MISO networks. Ni et al. \cite{10} presented a federated learning model designed to tackle the concurrent optimization challenges of beamforming and RIS phase reflection. This method enables parallel model training across multiple terminal devices and aggregates learned update information into a global model, enhancing data privacy protection and system optimization efficiency, which is particularly suitable for large-scale wireless networks. Moreover, He \textit{et al.}~\cite{10200991} proposed a convolutional neural networks (CNN) based framework to solve beamforming and RIS phase optimization problems in large-scale communication systems. By learning the spatial features of the wireless environment, CNN models can accurately predict optimal beamforming vectors and RIS phase configurations, significantly improving communication efficiency and quality.  Traditional DNNs and CNNs have limitations in handling non-Euclidean data. In contrast, GNN directly models the interactions of nodes among users, RISs, and relays through the message passing mechanism, and can dynamically adapt to channel changes and user grouping structure. For example, in a RIS-assisted multiuser MISO system, the GNN can map base stations, RISs, and relays as nodes in the graph, and the channel states are modeled as edge weights to jointly optimize beamforming and phase offset. This graph structure learning not only improves the fine-grained resource allocation, but also reduces the computational complexity through local information aggregation and overcomes the high computational overhead problem of traditional convex optimization methods.

Unsupervised learning methods have also been explored due to their potential to optimize RIS-aided systems without the need for labeled data~\cite{9427148}. These methods are particularly useful in scenarios where obtaining accurate CSI is challenging~\cite{12,13,14}. For instance, Song \textit{et al.}~\cite{12} designed an unsupervised learning framework to jointly optimize beamforming and reflection phases in RIS-assisted communication systems has been introduced. By utilizing CSI as inputs, this method can learn optimal beamforming strategies and RIS configurations without the need for explicit supervision signals. The concept has been extended to multi-user environments by integrating active and passive beamforming strategies within a dual-layer neural network framework~\cite{14}. Comparable methodologies leveraging CSI as inputs have been tailored to diverse scenarios with distinct model configurations~\cite{15}. Additionally, reinforcement learning (RL) has emerged as a potent approach for optimizing sum rate and energy efficiency in RIS-assisted wireless communication systems. The dynamic nature of RL makes it well-suited for addressing the complex and time-varying characteristics of wireless channels. Abdallah \textit{et al.}~\cite{10060056} crafted a multi-agent deep reinforcement learning model designed to concurrently optimize active beamforming at the base station and reflected beamforming by the RIS, utilizing solely received power measurements for this purpose. Xu \textit{et al.}~\cite{9779399} solved the beamforming problem in RIS-assisted millimeter-wave MIMO systems by optimizing the relevant parameters in the beamforming design and using location-aware mimicry environment and deep reinforcement learning (DRL) algorithms. Wang \textit{et al.}~\cite{9869783} proposed an algorithm based on deep Q-network (DQN). By discretization of trajectories, RIS can be used to assist unmanned aerial vehicle (UAV) communication systems to improve the communication quality between UAVs and UE. Yang \textit{et al.}~\cite{9322615} employed reinforcement learning algorithms to attain concurrent optimization of beamforming and RIS phase tuning. By designing a reward function to quantify communication performance, this algorithm autonomously learns how to adjust beamforming and phase configurations to enhance overall system performance.

\begin{table*}
\centering
\caption{Notations and their definitions.}
\label{tab:notation}
\begin{tabular}{c|p{7cm}|c|p{7cm}}
\bottomrule
Notations & Definitions & Notations & Definitions \\
\hline
$M$& Number of antennas at the BS &$N$ & Number of RIS elements\\
\hline
$L$& Number of antennas at the DF &$K$ & Number of users\\
\hline
$J$& Number of DF &$I$ & Number of groups\\
\hline
$P$& Transmit power &$\gamma_{\mathrm{R}}^{\mathrm{th}}$ & The threshold of DF\\
\hline
$\sigma_{k}^{2}$& Additive white Gaussian-noise of users &$\sigma_{\mathrm{R}}^{2}$
 & Additive white Gaussian-noise of DF \\
\hline
$G_i$&  The channel from the BS to the $RIS_i$ &$H_j^B$ & The channel from the BS to the $DF_j$\\
\hline
$H_{i,k}^B$&  The directed channel from the BS to user $k$ of $Group_i$ &$H_{i,k}^R$ & The channel from the $RIS_i$ to user k\\
\hline
$h_{i,j}$& The channel between the $RIS_i$ and the
$DF_j$ &$h_{j,k}^r$ & The channel from the $DF_j$ to user $k$\\
\hline
$y_k$& The BS transmits to user $k$ &$H_i$ & The cascaded channel between BS and user $k$ by $RIS_i$\\
\hline
$\mathbf{\theta}^i$& The reflection coefficient at $RIS_i$ &$x$ & The transmit signal\\
\hline
$g_k$& BS beamforming vector &$s_k$ & The source signal\\
\hline
$\gamma_{i,k}$& The SINR for user $k$ of $Group_i$ &$D$ & The update layers\\
\hline
$y_R$& The relay decodes signal of each user $k$ &$R_{i,j}$ &  The cascaded channel between BS and $DF_j$ by $RIS_i$\\
\hline
$\gamma_{j,k}^R$&  The SINR corresponding to user $k$ after filtering by $DF_j$ &$\alpha_k$ &  The combining filter for user $k$ at the relay\\
\hline
$x_R$&  The relay conveys signal & \( f_k \) & Relay beamforming vector\\
\hline
$C_{j,k}^i$&  The cascaded channel between $DF_j$ and user $k$ by $RIS_i$ &$R_{i,k}$ &  The sum rate of $Group_i$\\
\hline
$R_{i,k}^\mathbf{th}$&  The minimum rate threshold for user $k$ of $Group_i$ & $R_{i}^\mathbf{th}$ &  The minimum rate threshold for each $Group_i$\\
\toprule
\end{tabular}
\vspace{-0.1in}
\label{notation}
\end{table*}

However, existing works predominantly focus on optimizing the deployment and configuration of RISs without considering the diverse spatial distribution of users and their varying data rate demands, limiting the effectiveness of RIS optimization strategies in practical scenarios where users are distributed non-uniformly across the coverage area and exhibit varying communication needs. To address these issues, we propose an RIS- and DF-assisted MISO system with grouped users and fine-grained rate demands and introduce a two-phase GNN to facilitate the exchange and update of relational data within the graph-based model of a conjunct RIS- and DF-assisted multi-user MISO system, thereby acquiring efficient joint beamforming strategies through channel information extraction.

\section{Problem Formulation}
\label{system_model} 
In this paper, we consider a multiple RIS- and relay-assisted MISO system with multiple grouped users in the context of a large communication environment, as shown in Fig.~\ref{model}. 
Each user group is equipped with one RIS, and multiple user groups share a set of relays. By associating the RIS reflector with specific user groups, interference can be reduced and customized services can be achieved. This approach also enhances system flexibility by allowing for configurable relationships between RIS reflectors and user groups based on different communication scenarios and network topologies. Additionally, the shared use of relays optimizes resource utilization, avoids redundant deployment, improves system performance, and reduces construction and operation costs \cite{5740907,6570480}. We also consider different user groups with significant differences in geographic distribution. $Group 1$ is densely populated with users requiring low latency, while $Group 2$ has dispersed users requiring high throughput. The GNN models intra-group collaboration and inter-group competition through a graph structure, user nodes within a group share RIS reflection resources to achieve collaboration, while inter-group competition is dynamically balanced through relay selection. This topology-based learning mechanism enables the GNN to adaptively adjust beamforming strategies while meeting fine-grained rate requirements. The major notations used in this paper are defined in Table~\ref{notation}. Lower case letters represent scalars. Lower case bold-faced letters represent column vectors. Upper case bold-faced letters represent matrices. The transpose and Hermitian transpose of matrices are denoted by \((\cdot)^{\top}\) and \((\cdot)^{\mathrm{H}}\), respectively. \(\mathcal{CN}(\cdot, \cdot)\) stands for a complex Gaussian distribution.
\vspace{-0.1in}
\subsection{System Model}
\begin{figure*}[ht]
\centering
\adjustbox{width=18.5cm,height=9.3cm,keepaspectratio}{\includegraphics{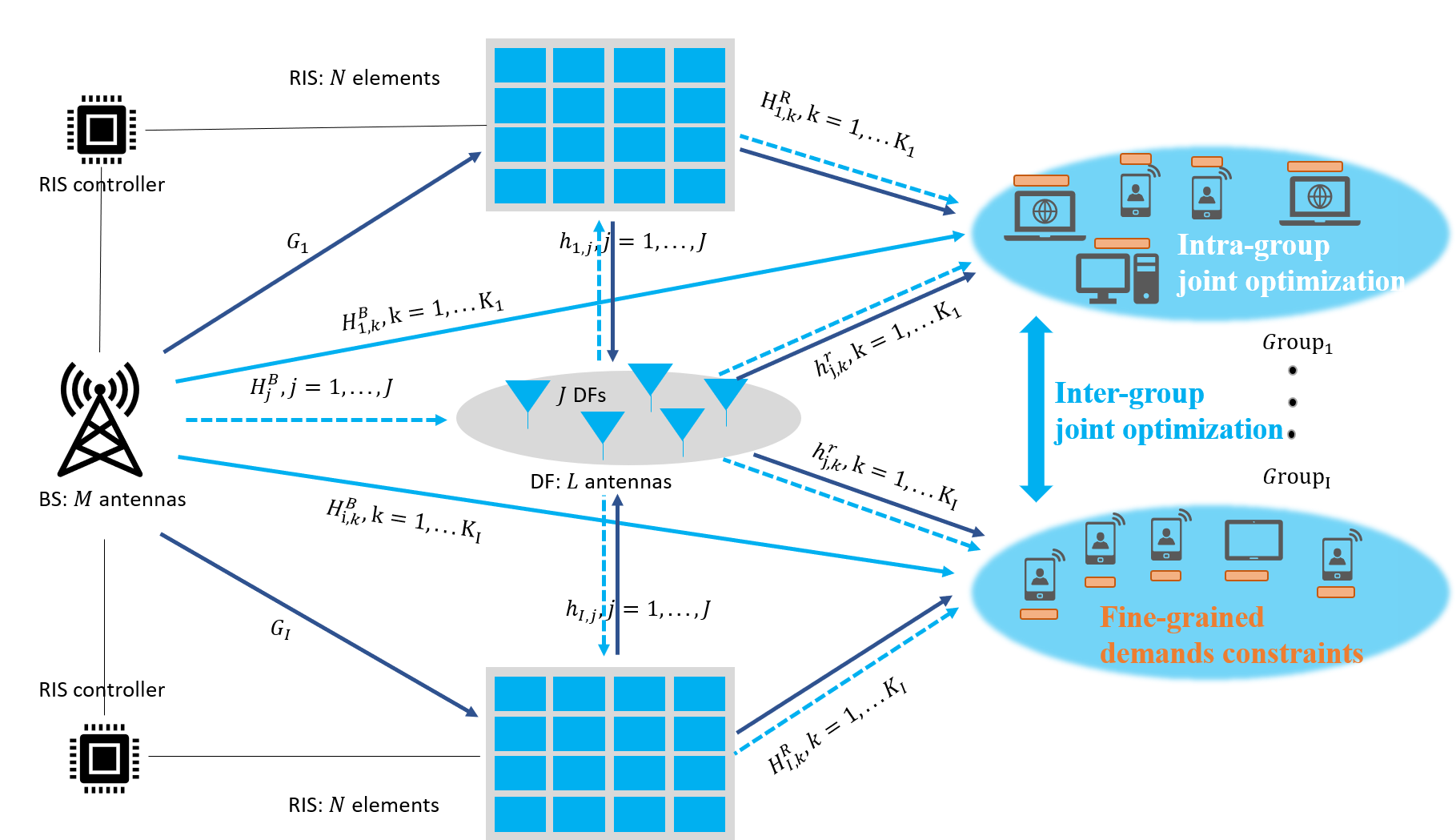}}
\vspace{-0.05in}
\caption{The proposed downlink communication for RIS-DF assisted MISO systems with multiple user groups, in which a group communicates through an RIS and a set of relays. Lines with different colors represent different paths.}
\label{model}
\vspace{-0.15in}
\end{figure*} 

As shown in Fig.~\ref{model}, we examine a two-phase multi-RIS-supported cooperative network, which includes a BS equipped with \( M \) antennas, \( I \) user groups with \( K \) single-antenna users per group, each group being served by an RIS comprising \( N \) elements, and \( J \) semi-duplex DF relays, each featuring \( L \) antennas. In half duplex mode, beamforming and phase offset strategies are designed according to the relayed frame protocol to maximize the throughput of the system and meet the user's rate requirements. Assuming full-duplex limited scenario, users need to transmit simultaneously to meet real-time requirements. Each RIS is equipped with a controller capable of adjusting the phases of array elements to reflect the incident signal in the desired direction. Furthermore, we assume that the channels from the BS to DF, BS to users, and DF to users are characterized as NLoS Rayleigh fading channels, while on the other hand, channels to and from the RIS are presumed to have LoS components modeled by Rician fading \cite{10041729,8746155,9235486}. In addition to the channels mentioned above, the DF relays play a crucial role in enhancing the communication between the BS and the users, which decode the signals received from the BS, re-encode and retransmit them to the users. This process helps in mitigating the effects of signal attenuation and improving the overall signal quality. The channels between the DF relays and the users are also modeled as NLoS Rayleigh fading channels, which are characterized by the presence of multipath propagation and significant signal scattering. The DF relays are strategically positioned to ensure effective coverage and reliable communication with the users, especially in areas where direct communication with the BS may be challenging due to obstacles or long distances. Hence, we denote the channel coefficient $h_{ab}$ between nodes a and b as:
\vspace{-0.03in}
\begin{equation}
\mathbf{h}_{ab}=\begin{cases}\bar{\mathbf{h}}_{ab},&\text{Rayleigh},\\
\sqrt{\frac{\kappa_{ab}}{\kappa_{ab}+1}}\hat{\mathbf{H}}_{ab}+\sqrt{\frac{1}{\kappa_{ab}+1}}\hat{\mathbf{h}}_{ab},&\text{Rician},\end{cases}
\end{equation}
where $\kappa_{ab}$ represents the Rician factor between nodes x and y. In a NLoS Rayleigh fading channel, $\begin{array}{rcl}\bar{\mathbf{h}}_{ab}&=&\bar{\mathbf{g}}_{ab}d_{ab}^{-\bar{\alpha}/2}\end{array}$, $x,y\in\{BS,DF,U\},x\neq y$, where $\overline{\mathbf{g}}_{ab}$ is modeled through zero-mean and unit-variance complex Gaussian small-scale fading, $d_{ab}$ represents the distance between nodes x and y, $\overline{\alpha}$ denotes the path loss exponent for the NLoS Rayleigh fading channel, and it is assumed that all channels remain constant without any change during the two phases. On the other hand, in the LoS Rician fading channel, $\hat{\mathbf{h}}_{ab}=\hat{\mathbf{g}}_{ab}d_{ab}^{-\hat{\alpha}/2}$, where $\hat{\alpha}$ denotes the path loss exponent for the LoS (Line-of-Sight) Rician fading channel, and $\hat{\mathbf{g}}_{ab}$ can be represented as: 
\vspace{-0.03in}
\begin{equation}
\hat{\mathbf{g}}_{ab}=\sqrt{\beta_{0}}[1,e^{-j\pi\sin\psi_{ab}},\ldots,e^{-j\pi(M-1)\sin\psi_{ab}}]^{T},
\end{equation}
where $\beta_{0}$ is the path loss at the reference distance $d_{0}=1m$, $\psi_{ab}$ is the angle of departure (AoD) or angle of arrival (AoA) of the signal between nodes a and b \cite{9235486}.

We denote the downlink channel matrix from the BS to the $i$-th RIS as $\mathbf{G}_i\in\mathbb{C}^{M\times N_i}$, denote the downlink channel matrix from the BS to the $j$-th DF as $\mathbf{H}_j^B\in\mathbb{C}^{M\times L_j}$, denote the directed downlink channel vectors from the BS to user $k$ of a different group as ${\mathbf{H}^B_{i,k}}\in \mathbb{C}^{M\times 1}$, denote the downlink channel vectors from the $i$-th RIS to the matched group's user $k$ as ${\mathbf{H}^R_{i,k}}\in \mathbb{C}^{N\times 1}$, denote the downlink channel matrix between the $i$-th RIS and the $j$-th DF as  $\mathbf{h}_{i,j}\in \mathbb{C}^{N_i\times L_j}$, and denote the downlink channel vectors from the $j$-th DF to user $k$ of a different group as ${\mathbf{h}^r_{j,k}}\in \mathbb{C}^{L\times 1}$. The communication process is divided into two phases. Initially, the BS sends signals to the users, with the signal received by user \( k \) expressed as:
\vspace{-0.03in}
\begin{align}
y_k^{(1)}&=(\mathbf{G}_idiag(\mathbf{\theta}_1^{i})\mathbf{H}^R_{i,k}+(\mathbf{H}^B_{i,k})^T)^T\mathbf{x}+\mathbf{n}_k^{(1)}  \notag\\
&=(\mathbf{H}_i\mathbf{\theta}_1^i+(\mathbf{H}^B_{i,k})^T)^T\mathbf{x}+\mathbf{n}_k^{(1)},
\label{channel_1}
\vspace{-0.03in}
\end{align} 
where \(\mathbf{H}_{i}=\mathbf{G}_{i}\text{diag}(\mathbf{H}^R_{i,k})\) represents the cascaded channel between the BS and user \(k\) through a \(\text{RIS}_{i}\), and \(\mathbf{n}_{k}^{(1)}\sim\mathcal{CN}(0,\sigma_{k}^2)\) represents the additive white Gaussian noise (AWGN). \(\mathbf{\theta}_{1}^{i}=[\mathbf{\theta}_{1,1},\ldots,\mathbf{\theta}_{1,N}]^{T}\) represents the reflection coefficient at \(\text{RIS}_{i}\) in the initial phase, the BS transmits the signal \(\mathbf{x}\), which is a sum of the source signals \(\mathbf{s}_{k}\) and the beamforming vectors \(\mathbf{g}_{k}\) for each user \(k\). This can be expressed as \(\mathbf{x} = \mathbf{G}\mathbf{s}\), where \(\mathbf{G}=[\mathbf{g}_1,\ldots,\mathbf{g}_K]\)  is the matrix of beamforming vectors and \(\mathbf{s}=[\mathbf{s}_1,\ldots,\mathbf{s}_K]^{\mathrm{T}}\) is the vector of source signals with \(\mathbb{E}[\mathbf{ss}^{\mathrm{H}}] = \mathbf{I}\). The signal-to-interference-plus-noise ratio (SINR) for user \( k \) is:
\vspace{-0.03in}
\begin{equation}
\gamma_{i,k}^{{(1)}}=\frac{\left|(\mathbf{H}_{i}\mathbf{\theta}_1^i+(\mathbf{H}^B_{i,k})^T)\mathbf{g}_{k}\right|^2}{\sum_{j=1,j\neq k}^{K}\left|(\mathbf{H}_{i}\mathbf{\theta}_1^i+(\mathbf{H}^B_{i,k})^T)\mathbf{g}_{j}\right|^2},
\end{equation}

In the subsequent stage, the relay transmits the decoded signals from the initial stage to users, assuming perfect decoding for each user \( k \) signal received in the first phase \cite{10041729}.
\begin{align}
y_{R}&=(\mathbf{h}_{i,j}^{H}{diag}({\mathbf{\theta}_1^{i}})\mathbf{G}_{i}+\mathbf{H}^B_{j})\mathbf{x}+{\mathbf{n}_{R}} \notag\\
& =(\mathbf{R}_{i,j}{\mathbf{\theta}_1^{i}}+\mathbf{H}^B_{j})\mathbf{x}+{\mathbf{n}_{R}}\label{channel_r},
\end{align}
where $\mathbf{R}_{i,j}=\mathbf{h}_{i,j}^{H}{diag}(\mathbf{G}_{i})$ represents the cascaded channel between the BS and $j$th relay through an $RIS_{i}$, and $\mathbf{n}_{R}{}\sim\mathcal{CN}(0,\sigma_{k}^2)$, considering the SINR for user \( k \) after the application of matched filter combining at the relay, i.e.,
\vspace{-0.03in}
\begin{equation}
\gamma_{j,k}^\text{R}=\frac{\|{\alpha}_k\|^4}{\sum_{j=1,j\neq k}^K{\alpha}_k^\text{H}{ \alpha }_j{\alpha}_j^\text{H}{ \alpha }_k+\sigma_{\text{R}}^2\|{\alpha}_k\|^2},
\vspace{-0.03in}
\end{equation}

The SINR for user \( k \) exceeds the threshold \(\gamma_{R}^{\text{th}}\) when using matched filter combining at the relay. The combining filter for user \( k \) is given by \(\alpha_k = (\mathbf{h}_{i,j}^{H} \text{diag}(\mathbf{\theta}_1^{i}) \mathbf{G}_{i} + \mathbf{H}^B_{j}) \mathbf{g}_{k}\). The relay then forwards the signal \(\mathbf{x}_R = \sum_{k=1}^{K} \mathbf{f}_{k} \mathbf{s}_{k} = \mathbf{F} \mathbf{s}\), where \(\mathbf{F} = [\mathbf{f}_1, \ldots, \mathbf{f}_K]\) represents the relay beamforming matrix, and \(\mathbf{s} = [\mathbf{s}_1, \ldots, \mathbf{s}_K]^{\mathrm{T}}\) is the vector of source signals with \(\mathbb{E}[\mathbf{ss}^{\mathrm{H}}] = \mathbf{I}\). The signal received by user \( k \) is:
\vspace{-0.03in}
\begin{align}
y_{k}^{({2})}&=\left({\mathbf{h}_{i,j}}{diag}({\mathbf{\theta}}_2^{i}){\mathbf{H}^R_{i,k}}+({\mathbf{h}^r_{j,k}})^T\right)^{{T}}\mathbf{x}_{{R}}+\mathbf{n}_{k}^{({2})}  \notag\\
&=\left(\mathbf{C}_{j,k}^{i}{\mathbf{\theta}}_2^{i}+({\mathbf{h}^r_{j,k}})^T\right)^{{T}}\mathbf{x}_{{R}}+\mathbf{n}_{k}^{({2})},
\label{eq7}
\vspace{-0.03in}
\end{align}
where $\mathbf{C}_{j,k}^{i}=\mathbf{h}_{i,j}{diag}(\mathbf{H}^R_{i,k})$ represents the cascaded channel between $DF_j$ and user $k$ through a $RIS_{i}$, and $\mathbf{n}_{k}^{{(}2{)}}\sim\mathcal{CN}(0,\sigma_{k}^2)$. $\mathbf{\theta}_{2}^i=[\mathbf{\theta}_{2,1},\ldots,\mathbf{\theta}_{2,N}]^{{T}}$ represents the reflection coefficient at $RIS_{i}$ in the second phase. The SINR for user \( k \) in the next stage is expressed as:
\vspace{-0.03in}
\begin{equation}
\gamma_{i,k}^{{(2)}}=\frac{\left|(\mathbf{C}_{j,k}^{i}{\mathbf{\theta}}_2^{i}+({\mathbf{h}^r_{j,k}})^T)\mathbf{f}_{k}\right|^2}{\sum_{j=1,j\neq k}^{K}\left|(\mathbf{C}_{j,k}^{i}{\mathbf{\theta}}_2^{i}+({\mathbf{h}^r_{j,k}})^T)\mathbf{f}_{j}\right|^2},
\vspace{-0.03in}
\end{equation}

Following the dual-phase transmission, the received signal at each user \( k \) is processed using maximum ratio combining (MRC), resulting in an overall SINR, i.e., $\gamma_{i,k}=\gamma_{i,k}^{\mathrm{(}1\mathrm{)}}+\gamma_{i,k}^{\mathrm{(}2\mathrm{)}}$. Thus, the sum rate from the BS to the user $k$ is as follows:
\begin{equation}
R_{i,k}=\log_2(1+\gamma_{i,k}).
\end{equation}

\vspace{-0.1in}
\subsection{Fine-grained Demands}\label{user}
We design an MISO system with grouped users where each user has a minimum rate threshold ${R_{i,k}^\mathrm{th}}$, which is termed the fine-grained rate demand. By satisfying a minimum rate threshold, we can ensure that each user can obtain at least enough data rates to meet their basic communication needs. This is critical to maintaining the user experience and meeting the underlying quality of service requirements. By personalizing rate settings for different users, network resources can be allocated more accurately, improving the overall network efficiency and resource utilization. 
\vspace{-0.05in}
\subsection{Problem Formulation}
The primary goal is to utilize the adopted GNN model to optimize the BS beamforming matrix, reflection coefficients phase shifts, and DF beamforming matrix. The optimization problem can be described as:
\vspace{-0.03in}
\begin{align}
\mathcal{P}_{1}: \;& \max_{\{\mathbf{g}_{k}\},\{\mathbf{f}_{k}\},\mathbf{\theta}_{1}^i, \mathbf{\theta}_{2}^i} \sum_{i=1}^{I}\sum_{k=1}^{K}R_{i,k}\label{10}\\ 
s.t.  \;\; & \mit{C_1} \text{:} \;\mathbb{E}[\parallel{\mathbf{x}}\parallel^2]=\mathrm{tr}(\mathbf{GG}^{\mathrm{H}})\leq {P}_{\mathrm{BS}}^{\mathrm{max}},\tag{10a}\label{10a}\\
& \mit{C_2} \text{:} \; \mathbb{E}[\sum_{j=1}^{J}\parallel{\mathbf{x}}_{{R}_j}\parallel^2]=\sum_{j=1}^{J}\mathrm{tr}\left(\mathbf{F}_j\mathbf{F}_j^{H}\right)\leq {P}_{R}^{max},\tag{10b}\label{10b}\\  
& \mit{C_3} \text{:} \; \gamma_{j,k}^\text{R}\geq\gamma_{\mathrm{th}}^{\mathrm{R}},\forall k=1,2,\ldots,K,\tag{10c}\label{10c}\\
& \mit{C_4} \text{:} \; R_{i,k}\geq{R_{i,k}^\mathrm{th}}, {R_{i}^\mathrm{th}},\forall k=1,2,\ldots,K,\tag{10d}\label{10d}\\
& \mit{C_5} \text{:} \; {\mathbf{\theta}}_{1}^i, {\mathbf{\theta}}_{2}^i\in{\mathbf{\theta}}^{N},\tag{10e}\label{10e}
\vspace{-0.03in}
\end{align}
where $C_1$ and $C_2$ represent the BS and relay transmission power constraints, respectively. $C_3$ ensures that the relay decoding is without issues. $C_4$ is the condition for group granularity and fine granularity requirements, and the user rate needs to exceed a minimum threshold. $C_5$ represents the RIS phase constraints, where $\begin{array}{rcl}\boldsymbol\Theta=\{e^{j\varphi_{n}}|\varphi_{n}\in\{0,\frac{2\pi}{2^B},\ldots,\frac{2\pi(2^B-1)}{2^B}\}\}\end{array}$, the finite resolution is B bits.
\vspace{-0.03in}
\section{Joint Optimization with Fine-grained Demands based on Two-phase GNN}
\label{method}
We introduce a GNN-driven approach for the communication system, where nodes represent network entities and possess features that encapsulate relational information. The GNN layers enhance data exchange and node updates, facilitating the simultaneous optimization of the base station's beamforming matrix \( G \), the relay's beamforming matrix \( F \), and the RIS phase shifts matrix \( \mathbf{\theta} \). This methodology is designed for two-stage transmissions within hybrid RIS and relay networks, with each stage utilizing a comparable GNN framework but with unique input data and graph configurations.
\vspace{-0.1in}
\subsection{The First Phase}
During the first phase, a fully connected graph is established, comprising \( K + 1 \) nodes, which lacks edge weights and is undirected. It consists of one node dedicated to the RIS for acquiring the phase shifts $\mathbf{\theta}_1^i$, along with $K$ nodes assigned to the users for the acquisition of their respective BS beamforming, i.e., $\mathbf{g}_{k},k=1,\ldots,K$. 

\subsubsection{Initial Layer}
The input encompasses the channel information related to the first-phase transmission from the BS to the relay and the uesr in the initial layer, as described in \eqref{channel_1} and \eqref{channel_r}, respectively. Particularly, the inputs are denoted as:
\vspace{-0.03in}
\begin{align}
\mathbf{H}_{k}^{(\mathrm{1})}&\triangleq[\mathbf{H}_i,(\mathbf{H}^B_{i,k})^T],k = 1,\ldots,K, \\
\mathbf{H_{R}}&\triangleq[\mathbf{R}_{i,j},\mathbf{H}^B_{j}],
\vspace{-0.05in}
\end{align}

We establish the initial feature vector \( r^{(0)} \) for the RIS nodes by leveraging the aggregated channel data between the users and the relay, encapsulated in \( \mathbf{H}_{k}^{(\mathrm{1})} \) and \( \mathbf{H_{R}} \). The feature extraction function $f^{(0)}:\mathbb{R}^{{2N\times{i}\times1}}\mapsto\mathbb{R}^{{q/2\times1}}$, capturing data from \( \mathbf{H}_{k}^{(\mathrm{1})} \), where \( q \) is a tunable parameter. Similarly, $\begin{aligned}f_\mathrm{R}^{(0)}:\mathbb{R}^{2ML\times{j}\times1}\mapsto\mathbb{R}^{q/2\times1}\end{aligned}$, extracting data from \( \mathbf{H_{R}} \). Initially, \( f^{(0)} \) is applied to \( \mathbf{H}_{k,m}^{(\mathrm{1})} \), which is \( m \)-th column of \( \mathbf{H}_{k}^{(\mathrm{1})} \). Subsequently, the mean operation over elements between \( k \) and \( m \) is employed to ensure permutation invariance in the GNN, denoted as \( \varphi_{\mathrm{mean}}(\cdot) \).
\begin{align}
\mathbf{r}_{k,m} & =f^{(0)}\big(\big[\mathbf{\Re}(\mathbf{H}_{k,m}^{\mathrm{(}1\mathrm{)}})^{\mathrm{T}},\mathbf{\Im}(\mathbf{H}_{k,m}^{\mathrm{(}1\mathrm{)}})^{\mathrm{T}}\big]^{\mathrm{T}}\big)\in\mathbb{R}^{q/2\times1},\forall m,\\ 
\mathbf{r} & =\varphi_{\mathrm{mean}}(\{\mathbf{r}_{k,m}\}_{k=1,\ldots,K,m=1,\ldots,M}\big)\in\mathbb{R}^{q/2\times1},
\vspace{-0.03in}
\end{align}
where $\mathbf{\Re}$ and $\mathbf{\Im}$ represent the real part and the imaginary part respectively. The initial features of RIS nodes are are acquired through combining the features extracted from \(\mathbf{H}_{k}^{(\mathrm{1})}\) using \(f^{(0)}\) and from \(\mathbf{H_{R}}\) using \(f_\mathrm{R}^{(0)}\), i.e.,
\vspace{-0.03in}
\begin{equation}
\mathbf{r}^{(0)}=\left[\mathbf{r}^{\mathrm{T}},f_{\mathrm{R}}^{(0)}\big(\big[\mathbf{\Re}(\mathrm{vec}(\mathbf{H}_{\mathrm{R}})),\mathbf{\Im}(\mathrm{vec}(\mathbf{H}_{\mathrm{R}}))\big]\big)^{\mathrm{T}}\right]^{\mathrm{T}}\in\mathbb{R}^{q\times1},
\vspace{-0.03in}
\end{equation}

The initial feature set for the \( k \)-th user node \( u_{k}^{(0)} \) is derived from \( \mathbf{H}_{k}^{(\mathrm{1})} \), using the extraction function $f_{\mathrm{u}}^{(0)}:\mathbb{R}^{2M\left(N\times{i}+1\right)}\mapsto\mathbb{R}^{q/2\times1}$, i.e.,
\vspace{-0.03in}
\begin{equation}
\mathbf{u}_k^{(0)}=f_u^{(0)}(\left[\mathbf{\Re}(\mathrm{vec}(\mathbf{H}_k^{(\mathrm{1})}),\mathbf{\Im}(\mathrm{vec}(\mathbf{H}_k^{(\mathrm{1})}))\right]^{\mathrm{T}})\in\mathbb{R}^{q\times1},
\vspace{-0.03in}
\end{equation}

\subsubsection{Node Update Layers}
Within the update layers, node features are updated through information aggregation and combination with neighbor nodes. The parameter \( D \), which signifies the number of update layers, is adjustable and can be configured as needed. During the \( d \)-th layer (\( d = 1, \ldots, D \)), the RIS node features are updated according to:
\begin{align}
\mathbf{r}^{(d)}&=\left[f^{(d)}\big(\big[\mathbf{r}^{(d-1)},\varphi_{\mathrm{mean}}(\big\{\mathbf{u}_{k}^{(d-1)}\big\}_{k=1,\ldots,K}\big)\big]\big),\mathbf{r}^{(d-1)}\right]  \notag \\
&\in\mathbb{R}^{q\left(d+1\right)\times1},
\vspace{-0.03in}
\end{align} 
where $f^{(d)}:\mathbb{R}^{2iqd\times1}\mapsto\mathbb{R}^{q\times1}$ is a node update function at the $d$th layer. Utilizing an element-wise averaging mechanism, each RIS node is allocated an equivalent volume of data from all user node. Moreover, to preserve the information from the preceding layer, the features are merged via concatenation. At the same time, the \( k \)-th user node features are modified by:
\vspace{-0.03in}
\begin{align}
\mathbf{u}_k^{(d)}&=\left[f_u^{(d)}(\big[\mathbf{u}_k^{(d-1)},\varphi_{\max}\big(\{\mathbf{u}_j^{(d-1)}\}_{\forall j\neq k}\big),\mathbf{r}^{(d-1)}\big]),\mathbf{u}_k^{(d-1)}\right] \notag
\\
&\in\mathbb{R}^{q\left(d+1\right)\times1},
\vspace{-0.05in}
\end{align} 

At each layer, the node update function $f_u^{(d)}:\mathbb{R}^{3iqd\times1}\mapsto\mathbb{R}^{q\times1}$, and the element-wise max function \( \varphi_{\max}(\cdot) \) is applied. This function ensures permutation invariance and enables each user node to identify maximum interference. The features extracted from the previous layer are incorporated to maintain historical data.

\subsubsection{Readout Layer}
Following the layers of node updates, the concluding features of the RIS and the user $k$ nodes are processed through the readout layer to determine the RIS phase shifts $\mathbf{\theta}_1^i$ and the BS beamforming $\mathbf{g}_k,{k=1,\ldots,K}$. We refer the readout layer as layer $D + 1$ for ease of notation. In the case of the RIS node, the output undergoes a linear function: 
\vspace{-0.03in}
\begin{equation}
\mathbf{r}^{(D+1)}=\left[r_{1}^{(D+1)},\ldots,r_{2N}^{(D+1)}\right]^{\mathrm{T}}=f^{(D+1)}(\mathbf{r}^{(D)})\in\mathbb{R}^{2N\times1},
\end{equation}
where ${f^{(D+1)}}:{\mathbb{R}^{q(D+1)\times1}}\mapsto\mathbb{R}^{2N\times1}$. The RIS phase shifts $\mathbf{\theta}_{1}=[\mathbf{\theta}_{1,n}]_{n=1,\ldots,N}$ are achieved from $r^{(D+1)}$ by initially calculating the continuous-phase $\widetilde{\mathbf{\theta}}_{1,n}^i$, i.e.,
\begin{equation}
\widetilde{\mathbf{\theta}}_{1,n}^i=\frac1{\sqrt{(r_n^{(D+1)})^2+(r_{N+n}^{(D+1)})^2}}\bigl(r_n^{(D+1)}+j\cdot r_{N+n}^{(D+1)}\bigr).
\end{equation}

Followed by the quantization of $\widetilde{\mathbf{\theta}}_{1,n}^i$ to the closest discrete phase in $\boldsymbol{\Theta}$. It should be emphasized that the continuous phase is utilized during training to enable backpropagation.

Similarly, final features of user nodes are processed by the readout layer to generate:
\begin{equation}
\left.\mathbf{u}_k^{(D+1)}=\left[\begin{matrix}u_{k,1}^{(D+1)},\ldots,u_{k,2M}^{(D+1)}\end{matrix}\right.\right]^{\mathrm{T}}=f_u^{(D+1)}(\mathbf{u}_k^{(D)})\in\mathbb{R}^{2M\times1},
\end{equation}
where $f_u^{(D+1)}:\mathbb{R}^{q(D+1)\times1}\mapsto\mathbb{R}^{2M\times1}$. The BS beamforming for user $k$ is contained as $g_{k,m}~=~u_{k,m}^{(D+1)}~+~j~\cdot~u_{k,M+m}^{(D+1)}$, which is normalized to satisfy the constraint of power in \eqref{10a}.

\begin{figure*}[ht]
\centering
\adjustbox{width=18cm,height=9cm,keepaspectratio}{\includegraphics{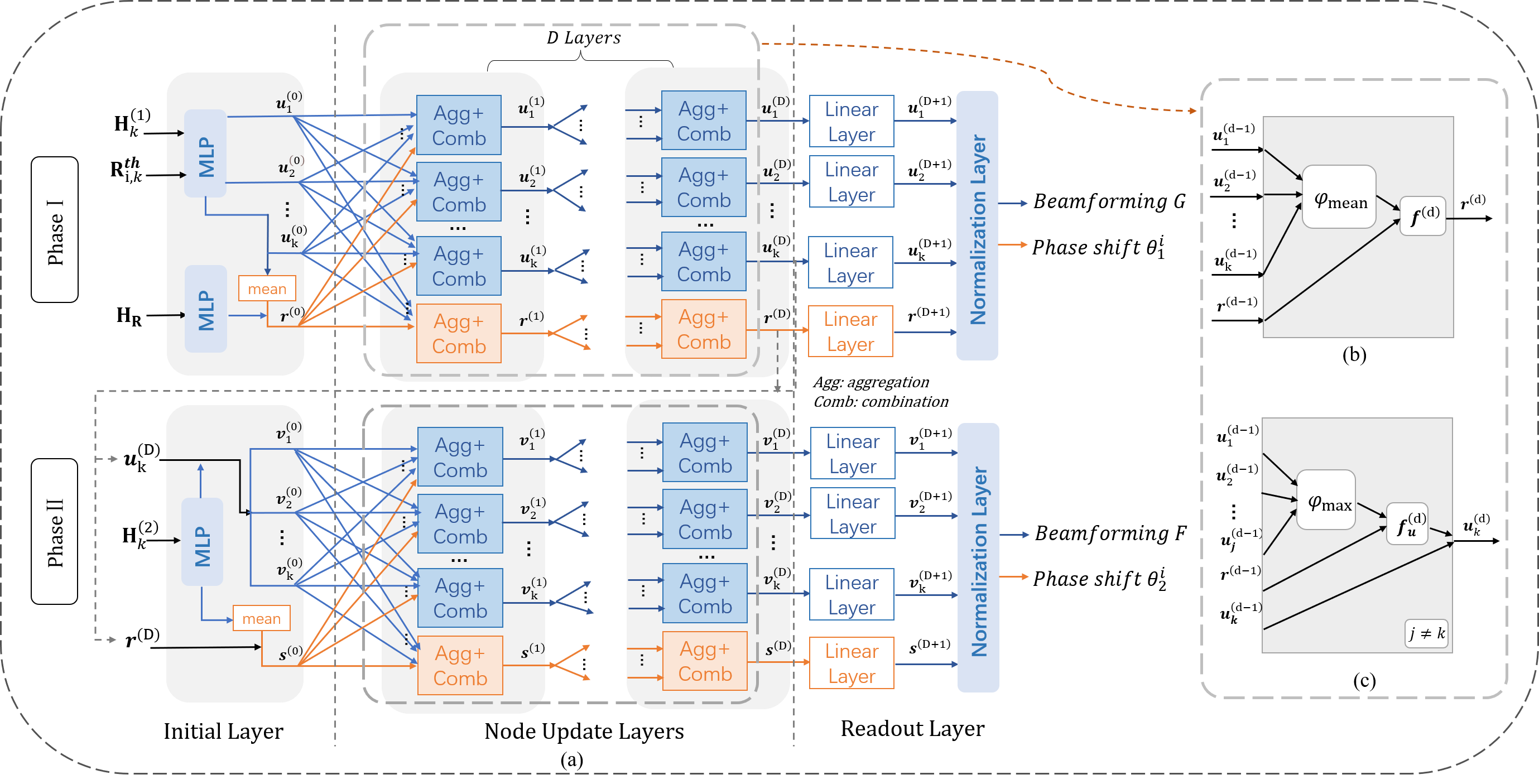}}
\caption{(a) The graph neural network architecture in different phases. Aggregation and combination operations of the $dth$ layer for (b) the IRS node, and (c) the user nodes.}
\label{phase}
\end{figure*} 
\vspace{-0.05in}
\subsection{The Second Phase}
In the subsequent phase of the GNN, the methodology is akin to the initial phase. Specifically, during this phase, the RIS node determines the phase shifts \(\mathbf{\theta}_2^i\), and each of the \(K\) user nodes calculates the relay beamforming \(f_k\) for \(k=1,\ldots,K\). The node features established in the first phase of the GNN serve as the basis for this subsequent phase.

\subsubsection{Initial Layer}
The initial features of RIS node incorporate the channel information for the transmission phase from the DF to the user, as specified \eqref{eq7}. The corresponding input is referred to as:
\vspace{-0.03in}
\begin{equation}
\mathbf{H}_{k}^{\mathrm{(2)}}\stackrel{\Delta}{=}[\mathbf{C}_{j,k}^{i},(\mathbf{h}_{j,k})^T],\;k=1,\ldots,K.
\vspace{-0.03in}
\end{equation}

The initial features of RIS node $s^{(0)}$ incorporate the aggregated channel information regarding the users and the relay, as encapsulated in $\mathbf{H}_{k}^{\mathrm{(2)}}$. Define feature extraction function \( g^{(0)} : \mathbb{R}^{2N \times i \times 1} \to \mathbb{R}^{q \times 1} \) to extract information from \(\mathbf{H}_{k}^{\mathrm{(2)}}\), where \( q \) is a tunable parameter. Firstly, $g^{(0)}$ is utilized on each column of $\mathbf{H}_{k}^{(\mathrm{2})}$, which is expressed as $\mathbf{H}_{k,\ell}^{(2)}$. Next is the average operation of elements between $k$ and $l$ to maintain the permutation invariance of the GNN, expressed as $\varphi_{\mathrm{mean}}(\cdot)$.
\vspace{-0.03in}
\begin{align}
\vspace{-0.03in}
\mathbf{s}_{k,\ell}&=g^{(0)}([\mathbf{\Re}(\mathbf{H}_{k,\ell}^{(2)})^{\mathrm{T}},\mathbf{\Im}(\mathbf{H}_{k,\ell}^{(2)})^{\mathrm{T}}]^{\mathrm{T}})\in\mathbb{R}^{q\times1},\forall\ell, \\
\mathbf{s}&=\varphi_{\mathrm{mean}}(\{\mathbf{s}_{k,\ell}\}_{k=1,\ldots,K,\ell=1,\ldots,L})\in\mathbb{R}^{q\times1}.
\end{align}

Additionally, in the second phase, the initial features of the RIS node are represented as:
\vspace{-0.03in}
\begin{equation}
\mathbf{s}^{(0)}=\left[\mathbf{s}^{\mathrm{T}},\left(\mathbf{r}^{(D)}\right)^{\mathrm{T}}\right]^{\mathrm{T}}\in\mathbb{R}^{q(D+2)\times1}.
\vspace{-0.03in}
\end{equation} 

The initial features for the \( k \)-th user node \( v_k^{(0)} \) are derived from \( \mathbf{H}_{k,\ell}^{(2)} \) using the feature extraction function $g_v^{(0)}:\mathbb{R}^{2L\times{j}(N+1)\times1}\mapsto\mathbb{R}^{q\times1}$, in conjunction with the final node features obtained from the initial phase of the GNN, i.e.,
\begin{align}
\mathbf{v}_{k}^{(0)}&=\Big[g_{v}^{(0)}\Big(\Big[\mathbf{\Re}(\mathrm{vec}(\mathbf{H}_{k}^{(\mathrm{II})})),\mathbf{\Im}(\mathrm{vec}(\mathbf{H}_{k}^{(\mathrm{II})}))\Big]^{\mathrm{T}}\Big),\mathbf{u}_{k}^{(D)}\Big] \notag\\
&\in\mathbb{R}^{q(D+2)\times1}.    
\end{align} 

\subsubsection{Node Update Layers}
At layer \( d \) (\( d = 1, \ldots, D \)), the features for both the RIS node and user nodes are updated accordingly through:
\begin{align}
\mathbf{s}^{(d)}&=\left[g^{(d)}\big(\big[\mathbf{s}^{(d-1)},\varphi_{\mathrm{mean}}\big(\{\mathbf{v}_{k}^{(d-1)}\}_{k=1,\dots,K}\big)\big]\big),\mathbf{s}^{(d-1)}\right]\nonumber \\
&\in\mathbb{R}^{q(D+d+2)\times1},\\
\mathbf{v}_{k}^{(d)}&=\left[g_{v}^{(d)}\Big(\big[\mathbf{v}_{k}^{(d-1)},\varphi_{\max}\big(\{\mathbf{v}_{j}^{(d-1)}\}_{\forall j\neq k}\big),\mathbf{s}^{(d-1)}\big]\Big),\mathbf{v}_{k}^{(d-1)}\right]\nonumber \\
&\in\mathbb{R}^{q(D+d+2)\times1}.
\vspace{-0.1in}
\end{align}

\subsubsection{Readout Layer}
Consistent with the function of the readout layer during the first phase, the relay beamforming $\mathbf{f}_k$ is derived from $\mathbf{v}_{k}^{(D+1)}=g_{v}^{(D+1)}(\mathbf{v}_{k}^{(D)})\in\mathbb{R}^{2L\times1},{k=1,\ldots,K}$, and the RIS phase shifts $\mathbf{\theta}_2^i$ are derived from $\mathbf{s}^{(D+1)}=g^{(D+1)}(\mathbf{s}^{(D)})\in\mathbb{R}^{2N\times1}$. Both are obtained by applying normalization and quantization to meet the conditions outlined in equations \eqref{10b} and \eqref{10c}. $g^{(D+1)}:\mathbb{R}^{2q(D+1)\times1}\mapsto\mathbb{R}^{2N\times1}$ and $g_{v}^{(D+1)}:\mathbb{R}^{2q(D+1)\times1}\mapsto\mathbb{R}^{2L\times1}$ are the layer functions.

The proposed framework is trained offline in an unsupervised manner, using diverse channel instances as training samples. We design loss function based on the objective in equation \eqref{10}, including an extra penalty term to address the constraint detailed in equation \eqref{10d} as follows:
\vspace{-0.05in}
\begin{equation}
\mathcal{L_C}=-\sum\limits_{i=1}^{I}\sum\limits_{k=1}^{K}R_{i,k}-\beta\sum\limits_{j=1}^{J}\sum\limits_{k=1}^{K}\min\left(0,\gamma_{j,k}^\text{R}-\gamma_{\text{th}}^{\text{R}}\right),
\label{loss1}
\vspace{-0.05in}
\end{equation}
where the weighting factor $\beta$ is ascertained based on empirical observation. The constraints in \eqref{10a}-\eqref{10d} are satisfied through the processes of normalization and quantization. 

The fine-grained division of group and user level allows for the selection of appropriate levels based on different environments. The group granularity involves establishing a minimum threshold rate for each group to ensure satisfying its total demand. The loss function is formulated as follows:
\begin{align}
\mathcal{L_G}=&-\sum\limits_{i=1}^{I}\sum\limits_{k=1}^{K}R_{i,k}-\beta\sum\limits_{j=1}^{J}\sum\limits_{k=1}^{K}\min\left(0,\gamma_{j,k}^\text{R}-\gamma_{\text{th}}^{\text{R}}\right) \notag\\
&-\lambda\sum\limits_{i=1}^{I}\sum\limits_{k=1}^{K}\min(0,R_{i,k}-R_{i}^{\text{th}}),
\label{loss2}
\end{align}
where $\mu$ is the weight coefficient of the penalty term, $R_{i}^{\text{th}}$ is the minimum group rate we set. In this manner, a penalty term can be incorporated into the loss function to impose a higher penalty for failing to satisfy group granularity demands, thereby catering to the fine-grained rate demands of different groups.  This approach allows for the flexible setting of fairness objectives and weights, balancing the rates based on different application scenarios.

Meanwhile, the fine granularity involves establishing a minimum threshold rate for each user to ensure satisfying users' fine-grained demands. The loss function is formulated as follows:
\begin{align}
\mathcal{L_F}=&-\sum\limits_{i=1}^{I}\sum\limits_{k=1}^{K}R_{i,k}-\beta\sum\limits_{j=1}^{J}\sum\limits_{k=1}^{K}\min\left(0,\gamma_{j,k}^\text{R}-\gamma_{\text{th}}^{\text{R}}\right) \notag\\
&-\lambda\sum\limits_{i=1}^{I}\sum\limits_{k=1}^{K}\min(0,R_{i,k}-R_{i,k}^{\text{th}}),
\label{loss3}
\end{align} 
where $\lambda$ is the weight coefficient of the penalty term, $R_{i,k}^{\text{th}}$ is the minimum user rate we set. Through a combination of normalization in the readout layer, penalty terms in the loss functions, and quantization operations, the constraints in $\mathcal{P}_{1}$ are effectively satisfied during the training and optimization process of the proposed two phase GNN based method.

\begin{algorithm}[ht!]
\caption{Joint Optimization with Fine-grained Demands based on Two-phase GNN(JOFD-TG) Algorithm}
\begin{algorithmic}[1]
\Require Initialize the parameters for First phase $\mathbf{H_k^{(1)}}$, $\mathbf{H_R}$, ${\mathbf{R}_{i,k}^\mathrm{th}}$; Second phase $\mathbf{H_k^{(2)}}$; 
\Ensure First phase ${\mathbf{\theta}}_1^i$, $\mathbf{G}$; Second phase ${\mathbf{\theta}}_2^i$, $\mathbf{F}$;  
\State \textbf{Initial:} $r^{(0)} \leftarrow \mathbf{H_k^{(1)}}, \mathbf{H_R};\enspace u_k^{(0)} \leftarrow \mathbf{H_k^{(1)}}, {\mathbf{R}_{i,k}^\mathrm{th}}$;
\For{$episode=1$ to $T$}
\State Small batch training samples are used;
\State Initialize $r^{(0)}$ and $u_k^{(0)}$; 
\For{$d=1$ to $D$}
\State $r^{(d)} \leftarrow COMB(AGG({u_k^{(d)}, k\in \mathcal{N}(k)}), r^{(d-1)});$
\State $u^{(d)} \leftarrow COMB(AGG({u_j^{(d-1)},\forall j \neq k},r^{(d-1)}),u_k^{(d-1)});$
\EndFor

\State Initialize $s^{(0)}$ and $v_k^{(0)}$;
\State \textbf{Initial2:} $s^{(0)} \leftarrow \mathbf{H_k^{(2)}}, r^{(D)};\enspace v_k^{(0)} \leftarrow \mathbf{H_k^{(2)}}, u_k^{(D)};$
\For{$d=1$ to $D$}
\State $s^{(d)} \leftarrow COMB(AGG({v_k^{(d)}, k\in \mathcal{N}(k)}), s^{(d-1)});$
\State $v^{(d)} \leftarrow COMB(AGG({v_j^{(d-1)},\forall j \neq k}, $
\State $\hspace{25 mm} s^{(d-1)}),v_k^{(d-1)});$   
\EndFor

\State According to (\ref{loss1})-(\ref{loss3}), calculating the loss $\mathcal{L}$ of batch samples, updating parameters using gradient descent. Select the appropriate relay to calculate the sum rate;
\EndFor
\State \textbf{Readout1:} ${\mathbf{\theta}}_1^i \leftarrow r^{(D+1)},\enspace \mathbf{G} \leftarrow u_k^{(D+1)};$
\State \textbf{Readout2:} ${\mathbf{\theta}}_2^i \leftarrow s^{(D+1)},\enspace \mathbf{F} \leftarrow v_k^{(D+1)};$
\end{algorithmic}
\end{algorithm}
\vspace{-0.1in}
\section{Performance Evaluation}
\label{result}
\subsection{Simulation Setting}
This section provides a numerical examination of the suggested algorithm. We focus on a multi-user MISO wireless communication setup, which is supported by two RISs, and two shared relays. The simulation parameters used for the analysis are shown in Table $\mathrm{\ref{simulation}}$. Following the setup described in \cite{9344820}, \cite{10184122}, we adopt the Adam optimizer to update the weights for network training, and set the initial learning rate to $0.001$, with a learning rate decay coefficient to $1e-6$, the batch size to $512$, and the number of hidden layer neurons to $128$ \cite{9427148,10184122}. The Rician factor $\kappa_{ab}$ between nodes is set to $10$ for LoS channels, consistent with typical urban micro-cell environments \cite{10200991}. The system features a BS equipped with $8$ antennas. Additionally, it includes four single-antenna users in each of the two considered groups, whose positions are randomly assigned within a circular area and are determined by the center of the circle \cite{10041729}. We model the following topology with different node distances (units are meters): the BS, RIS$_1$, RIS$_2$, DF$_1$ and DF$_2$ are positioned at coordinates $(0, 0)$, $(50, 100)$, $(50, -80)$, $(100, -10)$,  $(80, 25)$, respectively. Group1 users are scattered randomly within a circle of radius $10$ centered at coordinates $(200, 75)$, while Group2 users are distributed within a circle of the same radius centered at $(200, 10)$. The minimum user rate threshold $R_{i,k}^{\text{th}} = 1 \, \text{bps/Hz}$ is uniformly applied to all users to ensure fine-grained rate demands.

The channels for RIS-assisted and non-RIS-assisted systems are respectively characterized as quasi-static Rician and Rayleigh flat-fading models, as referenced in \cite{10041729} and \cite{9344820}. The functions $f_R^{(0)}$, $f^{(d)}$, $f_u^{(d)}$, $g^{(d)}$, and $g_v^{(d)}$ are realized using multi-layer perceptrons (MLP) with two suitable hidden layers for layers $d=0,1,\ldots,D$, and serve as linear transformations in the final layer $d=D+1$.
\begin{table}[ht!]
\vspace{-0.1in}
\renewcommand\arraystretch{1.25}
  \centering
  \caption{SIMULATION PARAMETERS}
    \begin{tabular}{| l | c |}
    \hline
    \textbf{Paramter} & \textbf{Value} \\
    \hline
    Number of antennas at the BS, $M$ & 8 \\
    \hline
    Number of DF, $J$ & 2 \\
    \hline
    Number of antennas at the DF, $L$ & 4 \\
    \hline
    Number of RIS elements, $N$ & 50 \\
    \hline
    Number of Groups, $I$ & 2 \\
    \hline
    Number of Users, $K$ & 4 \\
    \hline
    Transmit power, $P_{\mathrm{BS}}^{\mathrm{max}}= P_{\mathrm{R}}^{\mathrm{max}}$ & 20mW \\
    \hline
    Additive white Gaussian-noise, $\sigma_{k}^{2}=\sigma_{\mathrm{R}}^{2}$ & {2*10-5} \\
    \hline
    Threshold of DF, $\gamma_{\mathrm{R}}^{\mathrm{th}}$ & 0.01 \\
    \hline
    User rate threshold, $R_{i,k}^{th}$ & 1bps/Hz \\ 
    \hline
    Phase Shifts Parameter, $B$ & 2 \\
    \hline
    Number of node update layers, $D$ & 3 \\
    \hline
    {weighting factor, $\beta$} & 1000 \\
    \hline
    {weighting factor, $\lambda$} & 1000 \\
    \hline
    Adjustable parameter, $q$ & 128 \\    
    \hline
    Rician factor, $\kappa_{ab}$ & 10 \\
    \hline
    \end{tabular}%
\vspace{-0.1in}
  \label{simulation}%
\end{table}%

In the testing phase, our simulation refers to the proposed neural network model \cite{9621150,9427148,10184122} and the adopted benchmarking schemes are listed as follows:
\begin{itemize}
    \item \textbf{JOGD-TG}: the optimization of RIS and relay is performed jointly based on two-phase GNN, considering the threshold setting of groups, and maximizing the sum rate with group granularity.
    \item \textbf{JOCD-TG}: the optimization of RIS and relay is performed jointly based on two-phase GNN, regardless of the threshold setting, and aims to maximize the sum rate with coarse granularity. 
    \item \textbf{JOFD-DNN}: the optimization of RIS and relay is performed jointly based on DNN, considering the threshold setting of users, and maximizing the sum rate with fine granularity.   
    \item \textbf{JOFD-PSO}: a joint optimization scheme based on particle swarm optimization (PSO) searches for optimal beamforming and phase offset by iteratively updating the particle positions, and we uses pySwarms~\cite{pyswarmsJOSS2018} to implement the JOFD-PSO.
    \item \textbf{JOFD-Random}: random ${\mathbf{\theta}}_1^i$ and ${\mathbf{\theta}}_2^i$ without performing relay selection. Range reference Eqs.~\eqref{10e} for randomly generating ${\mathbf{\theta}}_1^i$ and ${\mathbf{\theta}}_2^i$.
\end{itemize}

\begin{figure}[ht!]
\centering
\subfloat[${\mathbf{\theta}}_1^i$ and ${\mathbf{\theta}}_2^i$]{\includegraphics[width=7.5cm,height=4.5cm]{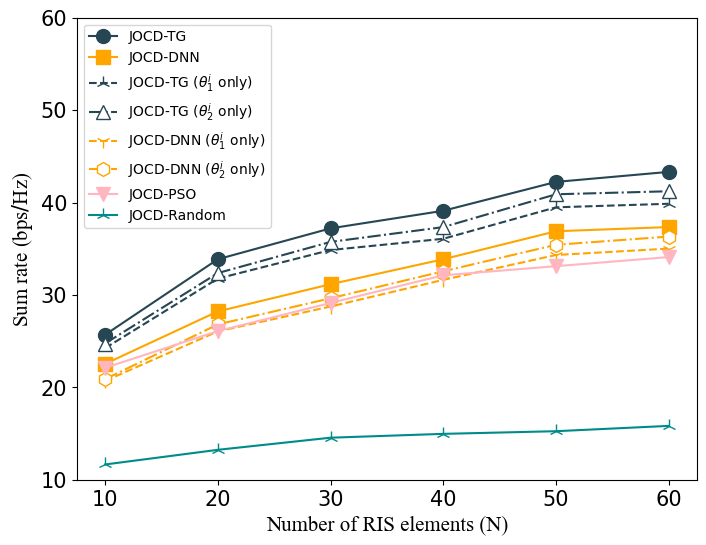} }
\\
\vspace{-0.1in}
\subfloat[Relay Selection]{\includegraphics[width=7.5cm,height=4.5cm]{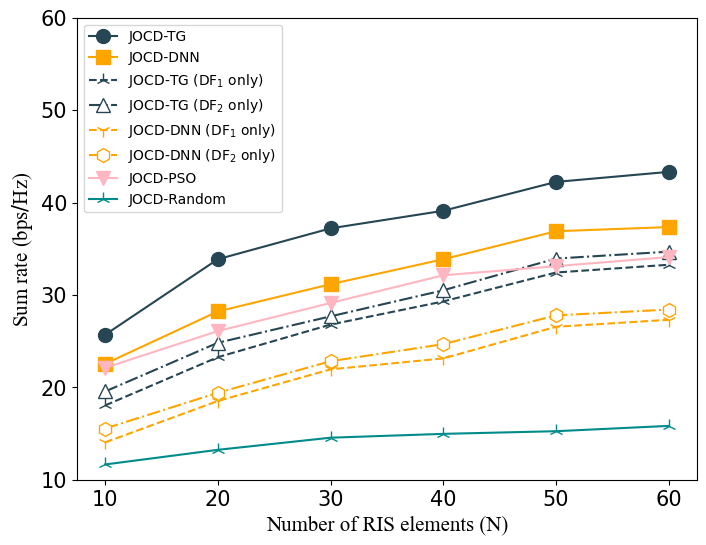}
}
\caption{The sum rate of different designs with or without ${\mathbf{\theta}}_1^i$, ${\mathbf{\theta}}_2^i$ and DFs}
\label{topo}
\end{figure}

\subsection{Experimental Results}

\subsubsection{Effect of the ${\mathbf{\theta}}_1^i$, ${\mathbf{\theta}}_2^i$ and relay selection}
 
Fig.~\ref{topo} depicts the comparison of the sum rate performance. In the mixed setup, the proposed JOCD-TG method, utilizing both ${\mathbf{\theta}}_1^i$ and ${\mathbf{\theta}}_2^i$, demonstrates superior performance compared to its simplified counterparts, and the performance of JOCD-TG surpasses that of JOCD-DNN. The variants employing only ${\mathbf{\theta}}_1^i$ or ${\mathbf{\theta}}_2^i$ exhibit slight performance degradation compared to the combined JOCD-TG scheme, highlighting the increased flexibility of JOCD-TG (${\mathbf{\theta}}_1^i$ and ${\mathbf{\theta}}_2^i$) in designing RIS phase shifts. The performance of JOFD-PSO in this scenario is slightly inferior to that of DNN, confirming its insufficient adaptability to high-dimensional optimization problems. In addition, GNN-based relay selection optimization is better than the case with a single relay. The appropriate relay can be selected for forwarding to enhance the final sum rate. Through joint optimization of RIS and relays, the impact of relay selection on system performance may be more significant than that of RIS phase tuning, attributing to the direct influence of relay positions and configurations on relay paths and attenuation, thus playing a more direct and prominent role in signal transmission. In contrast, the effect of RIS primarily manifests in signal reflection and phase adjustment, which have a relatively indirect impact on the signal. Therefore, in certain scenarios, relay selection may have a more direct influence on system performance compared to RIS phase tuning. 

\begin{figure}[h!]
\centering
\includegraphics[width=7.5cm,height=5.5cm]{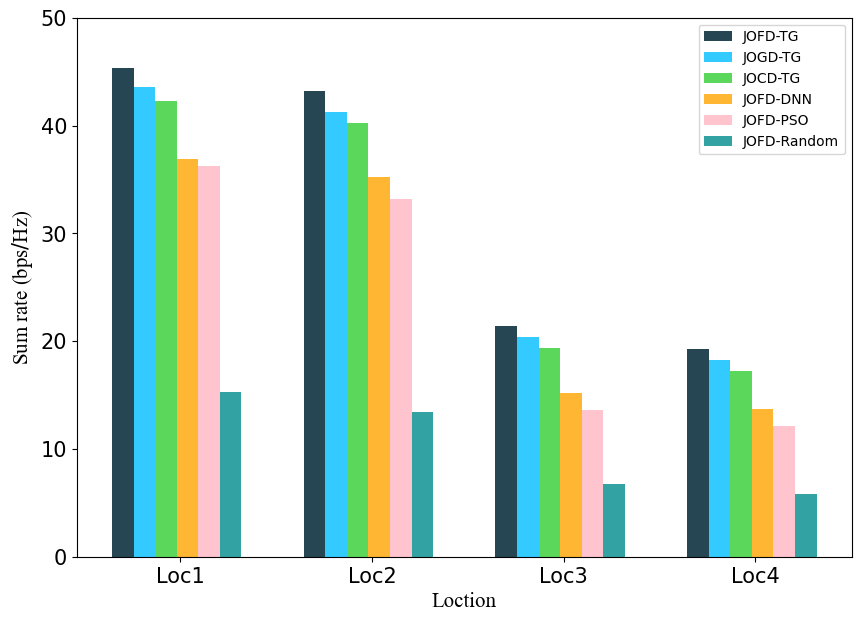}
\vspace{-0.1in}
\caption{The sum rate of different locations of RIS, relays and users}
\label{columnar}
\end{figure}

\subsubsection{Effect of the location of RISs and DFs}

Fig.~\ref{columnar} sets four different locations of the RIS$_1$, RIS$_2$, DF$_1$, DF$_2$ and the center of $Group1$ and $Group2$ with a radius of 10. We denote Loc1 to represents the positions set includes $(50, 100)$, $(50, -80)$, $(100, -10)$, $(80,25)$, $(200,75)$, and $(200,10)$, while the Loc2 set includes $(75, 100)$, $(75, -80)$, $(100, -10)$,  $(80, 25)$, $(200,75)$, $(200,10)$, the Loc3 set includes $(150, 100)$, $(150, -80)$, $(300, -10)$, $(240,25)$, $(600,75)$, $(600,10)$, and the Loc4 set includes $(225, 100)$, $(225, -80)$, $(300, -10)$, $(240, 25)$, $(600,75)$, $(600,10)$. When comparing different deployment locations, it was observed that RISs perform more effectively when located in proximity to the base station and relays are positioned closer to the user base. The proximity of RISs to the base station enables more efficient manipulation of the wireless signal, leading to optimized transmission to the user base. This closeness enables finer control over signal reflection, absorption, and redirection, resulting in improved signal quality and coverage. Additionally, reduced distance minimizes path losses and enhances the SINR, thereby enhancing overall system performance. Similarly, positioning the relays closer to the user base facilitates more efficient amplification and relay of signals with less attenuation. Proximity reduces transmission distance, path loss, and signal attenuation, resulting in a stronger received signal at the user terminal. Furthermore, the relays can operate at lower transmit power levels, reducing interference and improving signal clarity. JOFD-PSO is slightly worse and JOFG-Random performs the worst, highlighting the limitations of traditional heuristic algorithms and random strategies in multi-variable collaborative optimization scenarios.
 
As depicted in Fig.~\ref{columnar}, GNN consistently outperforms traditional algorithms when different sets of RIS, relay, and user groups are placed at various locations. The message-passing mechanism enables GNN to capture the dynamic topological information of local node neighborhoods, endowing it with strong generalization capabilities that allow for better adaptation to diverse environmental conditions compared to traditional algorithms. 

\begin{figure}[ht!]
\centering
\subfloat[ ($M, N, L, K_1, K_2$) = ($M$, 50, 4, 4, 4)]
{\includegraphics[width=7.5cm,height=4.5cm]{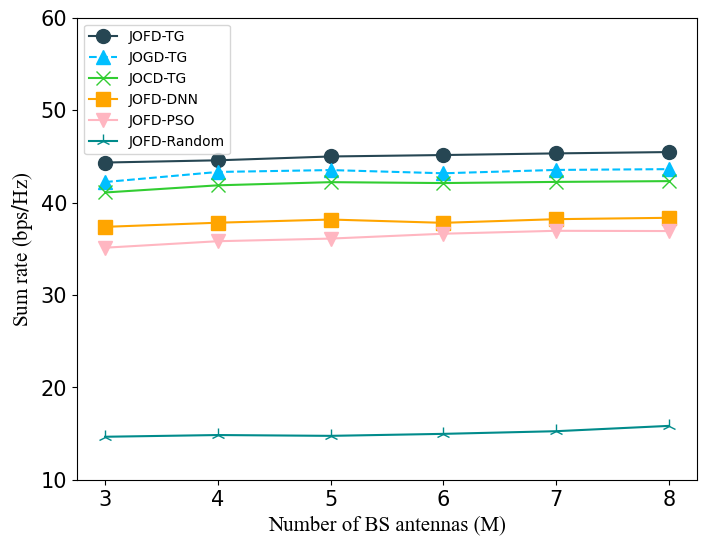}
\label{M_La}} 
\\
\subfloat[($M, N, L, K_1, K_2$) = (8, 50, $L$, 4, 4)]
{\includegraphics[width=7.5cm,height=4.5cm]{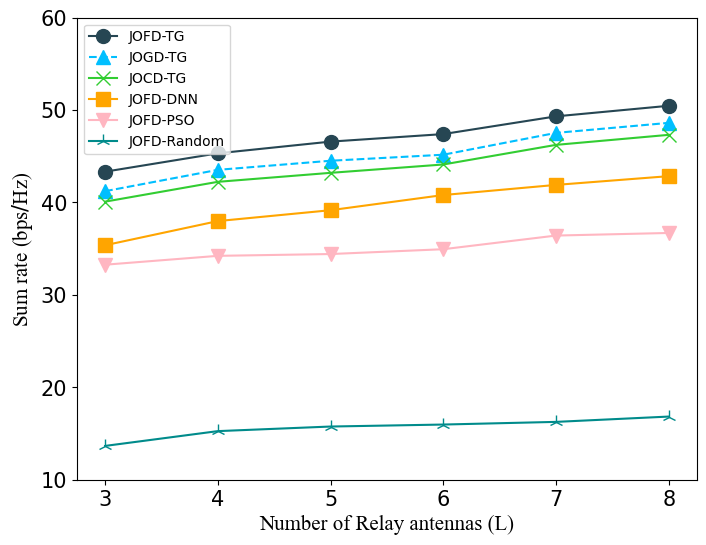}
\label{M_Lb}}
\caption{The sum rate of different numbers of BS antennas and relay antennas}
\label{M_L}
\end{figure}

\subsubsection{Effect of the number of BS antennas and relay antennas}
Fig.~\ref{M_La} shows the correlation between the sum rate and the quantity of base station antennas. It is observed that, across all assessed schemes, the effect of increasing \( M \) on the sum rate ranges from moderate to negligible. This is attributed to the topology of the network where the sum rate is primarily constrained by the second-stage transmission, characterized by a weak direct link. Consequently, enhancements in $M$ slightly improve the SINR for the first-stage end-user, yielding only marginal gains. 

Fig.~\ref{M_Lb} explores the correlation between the relay antenna count and the sum rate. The results indicate that as $L$ increases, the sum rate for all schemes shows an enhancement. This improvement stems from the increased degrees of freedom available during the second stage, allowing for more precise tuning of the relaying beamforming vector, which optimally supports the relay-assisted second-stage transmission to the end-user across auxiliary and direct joint channels. JOFD-TG, which incorporates intra-group cooperation via group-level rate thresholds, outperforms JOCD-TG by 15$\%$ in sum rate, highlighting the significance of intra-group resource balancing. JOFD-PSO still performs slightly worse in this case. Because it is limited by the heuristic search mechanism, its sum rate remains lower, which verifies the limitations of traditional optimization algorithms in complex interaction case.

\begin{figure}[h!]
\centering
\includegraphics[width=7.5cm,height=5.5cm]{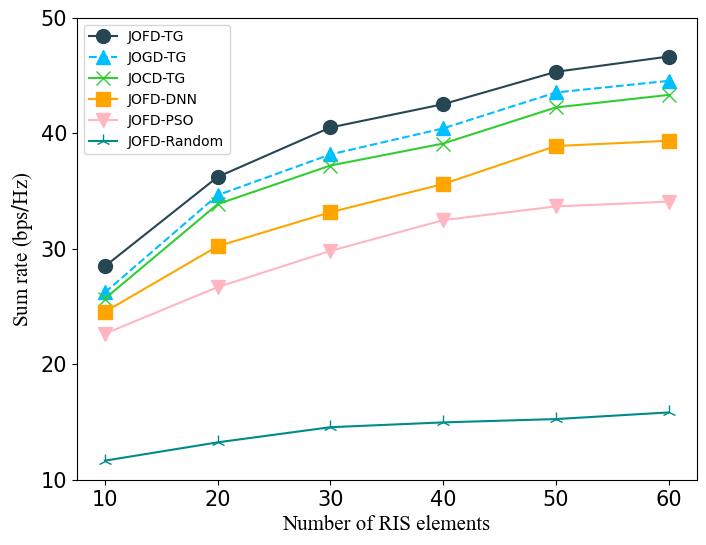}
\vspace{-0.1in}
\caption{The sum rate of different numbers of RIS elements}
\label{fine-sum}
\end{figure}

\begin{table}[ht!]
  \centering
  \caption{SUM RATE VS. TRAINING (ROW) AND TESTING (COLUMN) USER COUNTS $K$}
    \begin{tabular}{|c|c|c|c|}
    \hline
    Sum rate (bps/Hz) & $K = 20$ & $K = 25$ & $K = 30$ \\
    \hline
    Same $K$ as testing & 19.34 & 19.25 & 19.08 \\
    \hline
    $K = 10$ & 18.27(94.5\%) & 17.98(93.4\%) & 17.76(93.1\%) \\
    \hline
    $K = 15$ & 18.62(96.3\%) & 18.33(95.2\%) & 18.06(94.7\%) \\
    \hline
    $K = 20$ & 19.25(99.5\%) & 19.17(99.1\%) & 18.79(98.5\%) \\
    \hline
    \end{tabular}%
  \label{userk}%
\end{table}%

\subsubsection{Effect of the number of users}
Table \ref{userk} evaluates the proposed GNN model's ability to generalize across different user counts. It shows that the model, trained with a smaller \( K \), maintains acceptable performance when evaluated with a larger \( K \). This resilience is attributed to the user-independent feature extraction and permutation invariance of the model. Nonetheless, performance degrades as the discrepancy between training and testing \( K \) values grows. The GNN’s permutation-invariant design, as validated in Table~\ref{userk}, ensures robust generalization across varying user counts by extracting user-independent features. The offline unsupervised training, validated cross-K performance by Table~\ref{userk}, avoids the need for labeled data by directly optimizing system objectives. This aligns with practical scenarios where labeled data is scarce.

Previous experiments have focused on addressing the challenge of the rate maximization. As mentioned in Section \ref{user}, we divided the user fine granularity and each user has a minimum rate threshold. As depicted in Fig. \ref{fine-sum}, the sum rate of fine-grained granularity is superior to that of coarse-grained granularity, and the sum rate at the user level is better than at the group level. Fine-grained granularity excels primarily because it offers greater flexibility and precision, allowing for more detailed and personalized resource management and service customization. Our approach to designing thresholds for different users enhances adaptability and improves both the specificity and efficiency of services. In the meantime, our approach offers several advantages over alternative methods.

\subsubsection{Convergence Performance}
\begin{figure*}[ht!]
\centering
\subfloat[Sum Rate]
{\includegraphics[width=5.8cm,height=5cm]{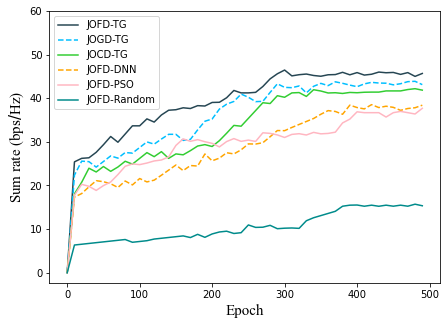}} 
\subfloat[Satisfaction rate]
{\includegraphics[width=5.8cm,height=4.9cm]{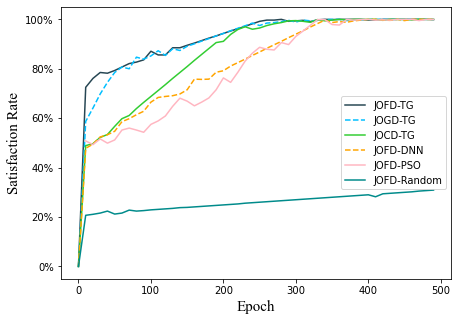}}
\subfloat[Loss]
{\includegraphics[width=5.8cm,height=4.9cm]{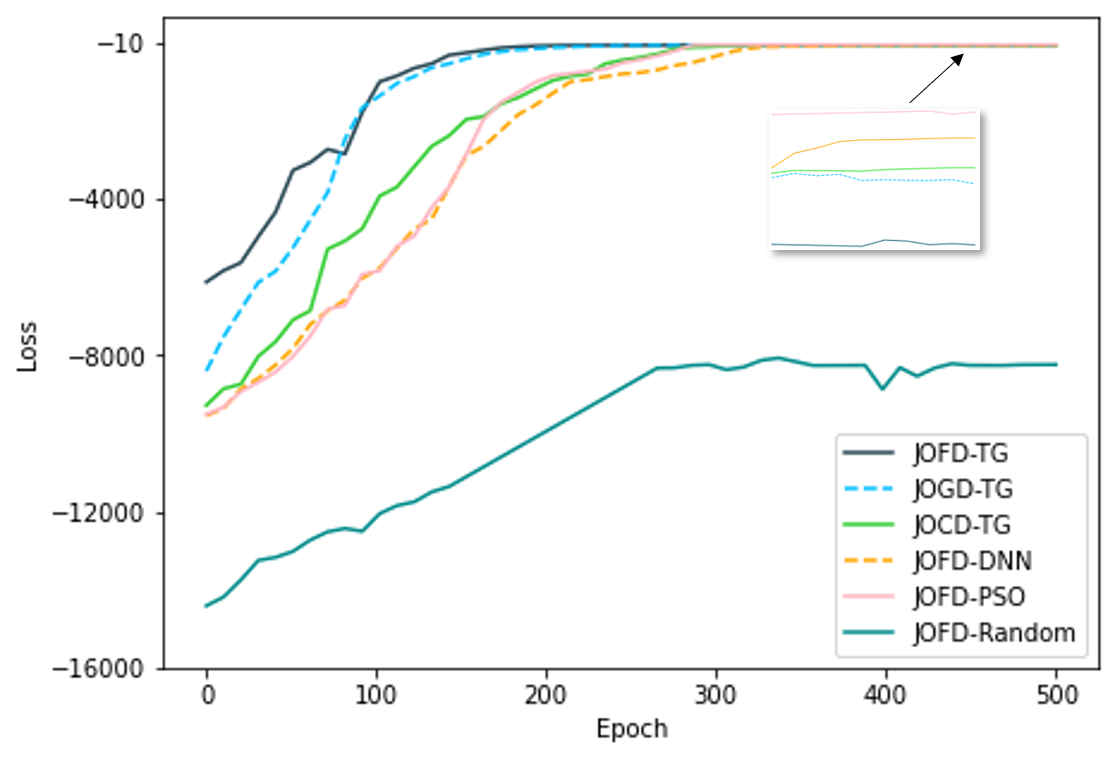}}
\caption{The sum rate, satisfaction rate and loss of different epochs}
\label{epoch}
\end{figure*}

Fig.~\ref{epoch} demonstrates the convergence of all models after numerous epochs. The convergence of the iterative optimization process signifies that further iterations do not yield significant changes in the output, indicating that the model has effectively learned and adapted to the training data, achieving a stable and satisfactory level of performance. with the JOFD-TG model consistently outperforming others in terms of maximum sum rate and satisfaction rate. In addition to JOFG-Random, it is noteworthy that the loss function value for JOFD-TG is the highest among all models, primarily due to the user's rate exceeding the threshold, resulting in negligible penalty term loss and a predominant influence from maximum user and rate.

\subsubsection{Effect of the weight coefficient $\lambda$}

\begin{figure}[ht!]
\centering
\subfloat[Sum Rate]
{\includegraphics[width=7.5cm,height=4.5cm]{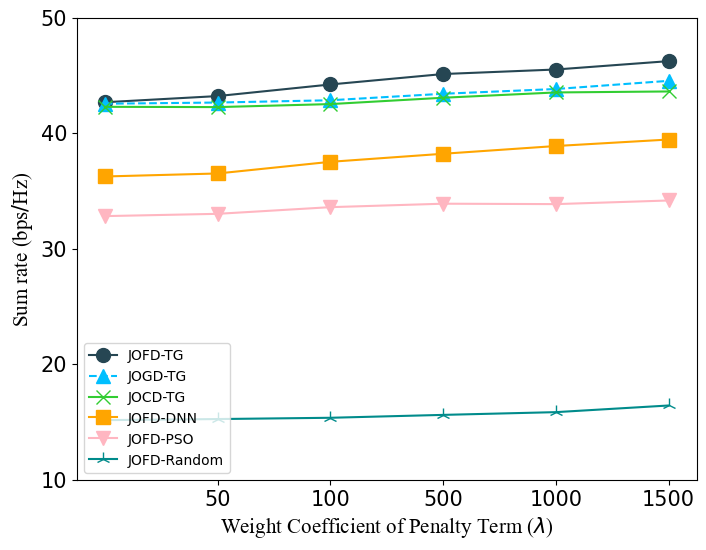}
\label{parameter_a}} 
\\
\subfloat[Satisfaction Rate]
{\includegraphics[width=7.9cm,height=4.5cm]{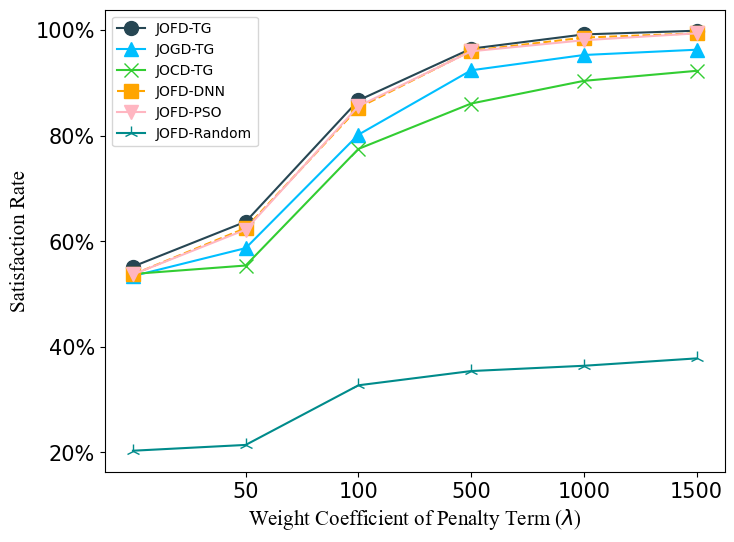}
\label{parameter_b}}
\caption{The sum rate and satisfaction rate of different weight coefficients $\lambda$}
\label{parameter}
\end{figure}

In Fig.~\ref{parameter_a}, we use varying minimum rate thresholds for different numbers of groups and individual users, and can observe the effects by adjusting the weight coefficient $\lambda$. After the weight coefficient exceeds 1000, the maximum sum rate of several models and the satisfaction rate of coarse-grained and fine-grained tend to stabilize. Then, when the weight coefficient is 2000, the satisfaction rate of JOFD-TG tends to be 100$\%$. It is noteworthy that JOFD-TG consistently outperforms other models in terms of the maximum sum rate of users and the percentage of users surpassing the threshold. Meanwhile, user-level fine granularity enables personalized resource allocations tailored to individual requests and usage patterns, whereas group-level granularity determines resource allocation based on average or collective needs. This user-level granularity facilitates rapid response to fluctuations in individual user needs, while fineness at the group level may exhibit limitations in responding to sudden changes due to considerations of resource allocation and balance across the entire group, potentially resulting in reduced overall performance.

In Fig.~\ref{parameter_b}, it is evident that as the weight coefficient of the penalty term increases, the satisfaction rate also increases in all models. Ultimately, the satisfaction rate based on the fine-grained model was close to 100 $\%$. This demonstrates the viability of our proposed fine-grained partitioning based on minimum thresholds in ensuring that model outputs adhere to threshold constraints. Furthermore, it provides further evidence for the adaptability and flexibility of the model to accommodate different performance requirements through parameter adjustments, thus showcasing its robustness. The experimental results validate that neglecting either inter-group competition or intra-group cooperation leads to suboptimal performance. JOCD-TG achieves lower satisfaction rate than JOFD-TG, while JOFD-DNN fails to generalize across diverse node deployments.

Regarding the experimental analysis of Fig.~\ref{parameter_b}, it further supports the effectiveness of our method, showing the changes in the total system rate and user rate requirement satisfaction rate under different weight coefficients $\lambda$. As $\lambda$ increases, the satisfaction rate gradually increases, which shows that by adjusting $\lambda$, we can effectively control the degree of constraint satisfaction. When $\lambda$ reaches 2000, the satisfaction rate of the JOFD-TG method approaches 100$\%$, which means that almost all user rate requirements are met. This proves that our loss function design can effectively ensure the satisfaction of constraints under appropriate parameter settings.

\section{Conclusion}
\label{conclusion}
In this paper, we proposed a novel joint optimization method for multiple RISs- and DFs-assisted MISO systems that serves to maximize the sum rate while meeting grouped user fine-grained demands. We designed a new loss function to accommodate the diverse demands of user groups by incorporating minimum thresholds. The proposed GNN model can be tailored to autonomously learn efficient phase shifts and beamforming directly from input CSI, while also performing simultaneous relay selection. Simulation results demonstrated the superior performance of this approach, as well as its scalability across varying numbers of users.

In the future work, we will explore the development of systems based on Stacked Intelligent Metasurfaces (SIM), incorporating one SIM at the transmitter and another at the receiver. Unlike conventional architectures, SIM enables direct precoding during transmission and merging during reception within the electromagnetic wave propagation process. Traditional communication systems necessitate a substantial number of RF links, while SIM technology facilitates signal processing directly in the electromagnetic domain, thereby substantially reducing the reliance on RF links~\cite{10158690}.

\section*{Acknowledgments}
This work was supported by the Fundamental Research Funds for the Zhejiang Provincial Natural Science Foundation of China under Grant (No. LQN25F020016 and LDT23F01012F01), National Natural Science Foundation of China (No. 62401190, 62071327 and 62372146), Tianjin Science and Technology Planning Project (No. 22ZYYYJC00020),  and Engineering and Physical Sciences Research Council grant (No. EP/X040518/1 and EP/Y037421/1).

\bibliographystyle{IEEEtran}
\bibliography{IEEEabrv,reference}

\begin{IEEEbiography}[{\includegraphics[width=1in,height=1.25in,clip,keepaspectratio]{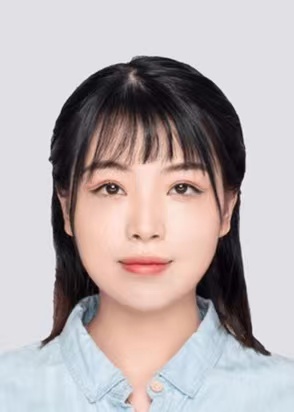}}]{Huijun Tang} received the BSc degree from Jinan University, China in 2016 and the M.S. and Ph.D. degree from Tianjin University, China in 2018 and 2022, respectively. She is currently a Lecturer with the School of Cyberspace, Hangzhou Dianzi University. Her research interests include internet of things, mobile edge computing, and complex networks.
\end{IEEEbiography}

\begin{IEEEbiography}[{\includegraphics[width=1in,height=1.25in,clip,keepaspectratio]{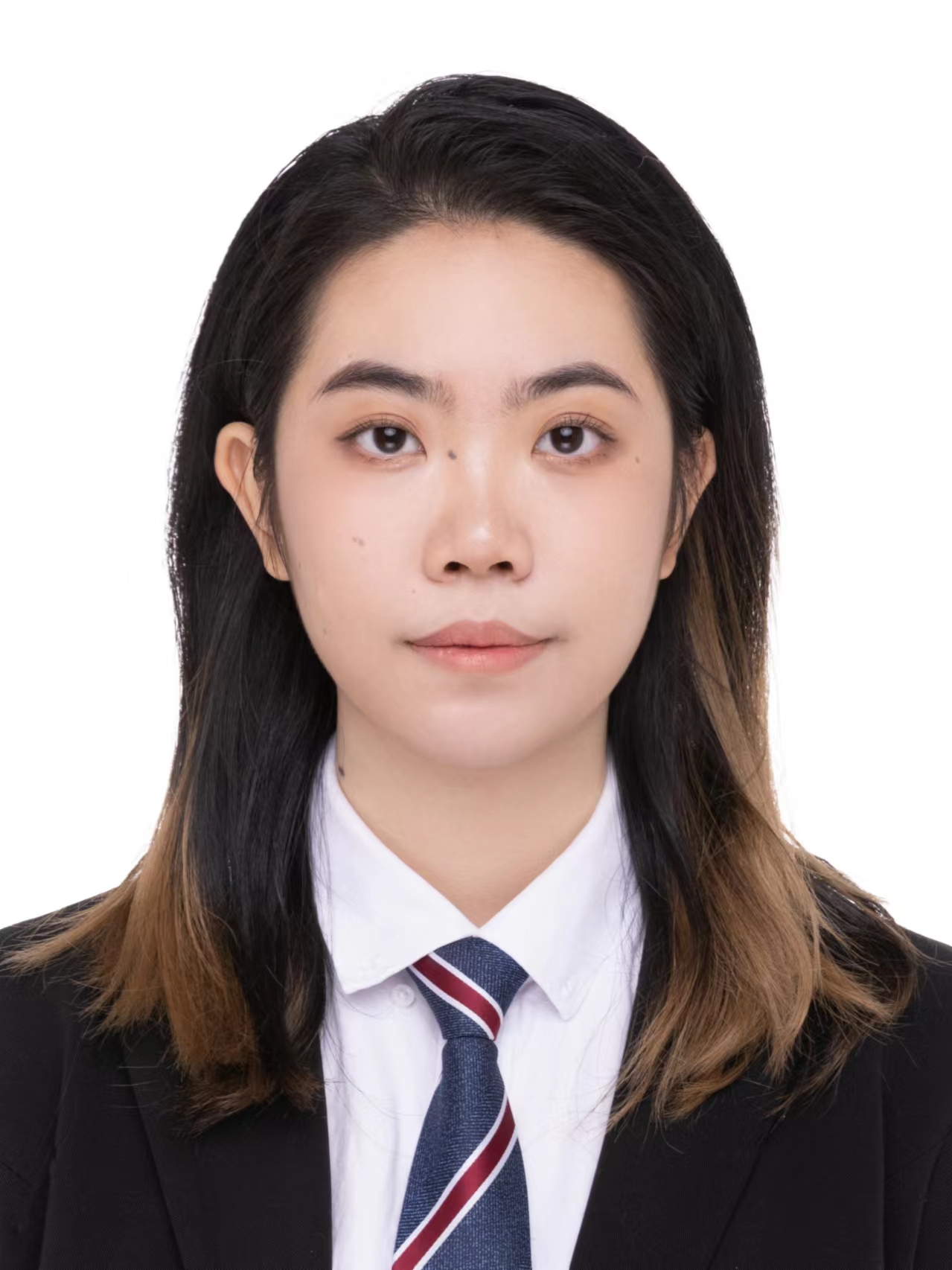}}]{Jieling Zhang} received the Bachelor degree from Hangzhou Dianzi University in 2023. She is currently pursuing a master's degree at the School of Cyberspace, Hangzhou Dianzi University. Her current research interests include signal processing and optimization and graph generation.
\end{IEEEbiography}

\begin{IEEEbiography}[{\includegraphics[width=1in,height=1.25in,clip,keepaspectratio]{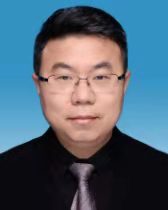}}]{Zhidong Zhao} received the B.S. and M.S. degrees in mechanical engineering from the Nanjing University of Science and Technology, China, in 1998 and 2001, respectively, and the Ph.D. degree in biomedical engineering from Zhejiang University, China, in 2004. He is currently a Full Professor at Hangzhou Dianzi University, China. His research interests include biomedical signal processing, wireless sensor networks, biometrics, and machine learning.
\end{IEEEbiography}

\begin{IEEEbiography}[{\includegraphics[width=1.0in,trim=0in 0.7in 0in 0.3in,clip,keepaspectratio]{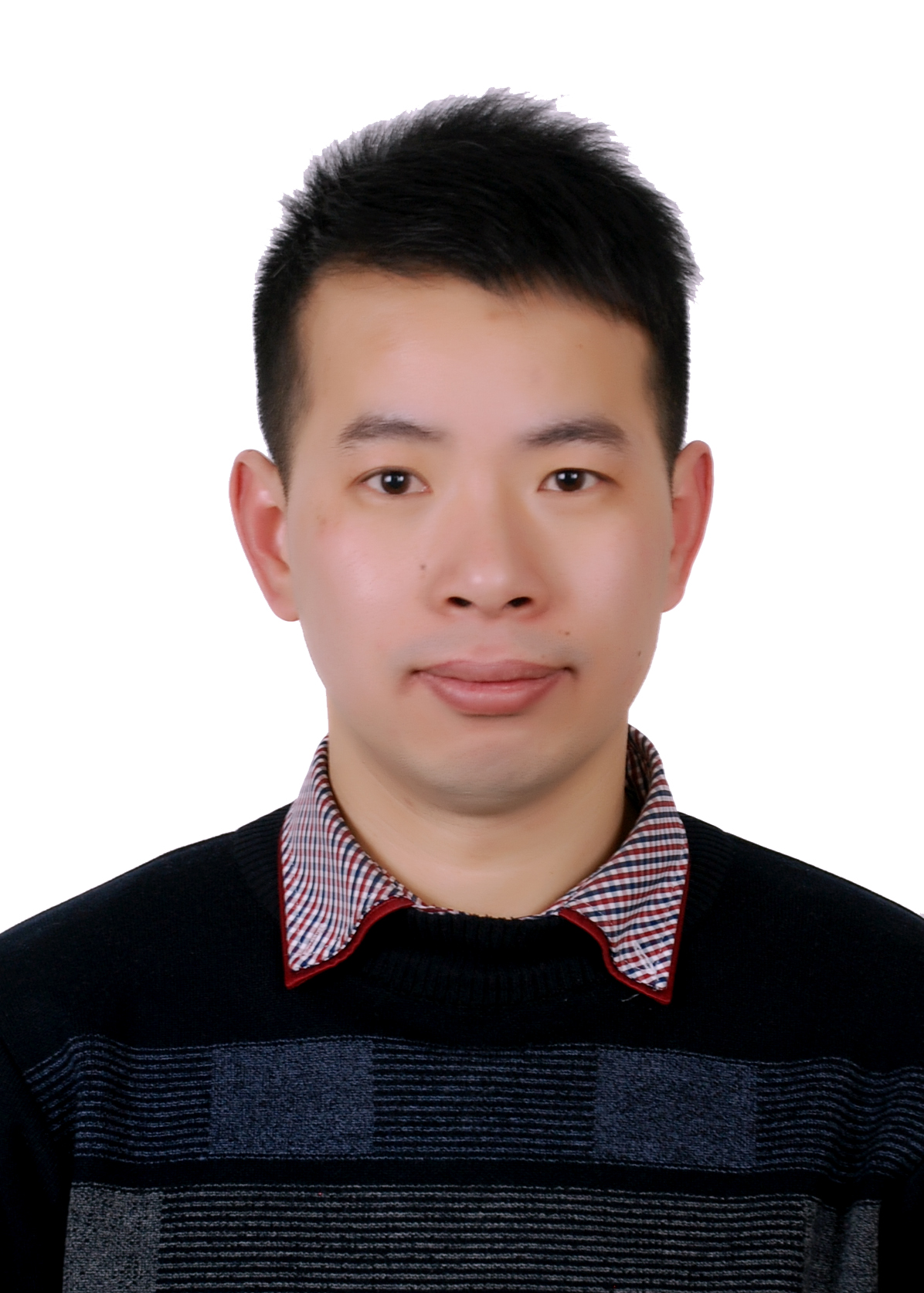}}]{Huaming Wu} (Senior Member, IEEE) received the B.E. and M.S. degrees from Harbin Institute of Technology, China in 2009 and 2011, respectively, both in electrical engineering. He received the Ph.D. degree with the highest honor in computer science at Freie Universit\"at Berlin, Germany in 2015. He is currently a professor at the Center for Applied Mathematics, Tianjin University, China. His research interests include mobile cloud computing, edge computing, internet of things, deep learning, complex networks, and DNA storage.
\end{IEEEbiography} 

\begin{IEEEbiography}[{\includegraphics[width=1in,height=1.25in,clip,keepaspectratio]{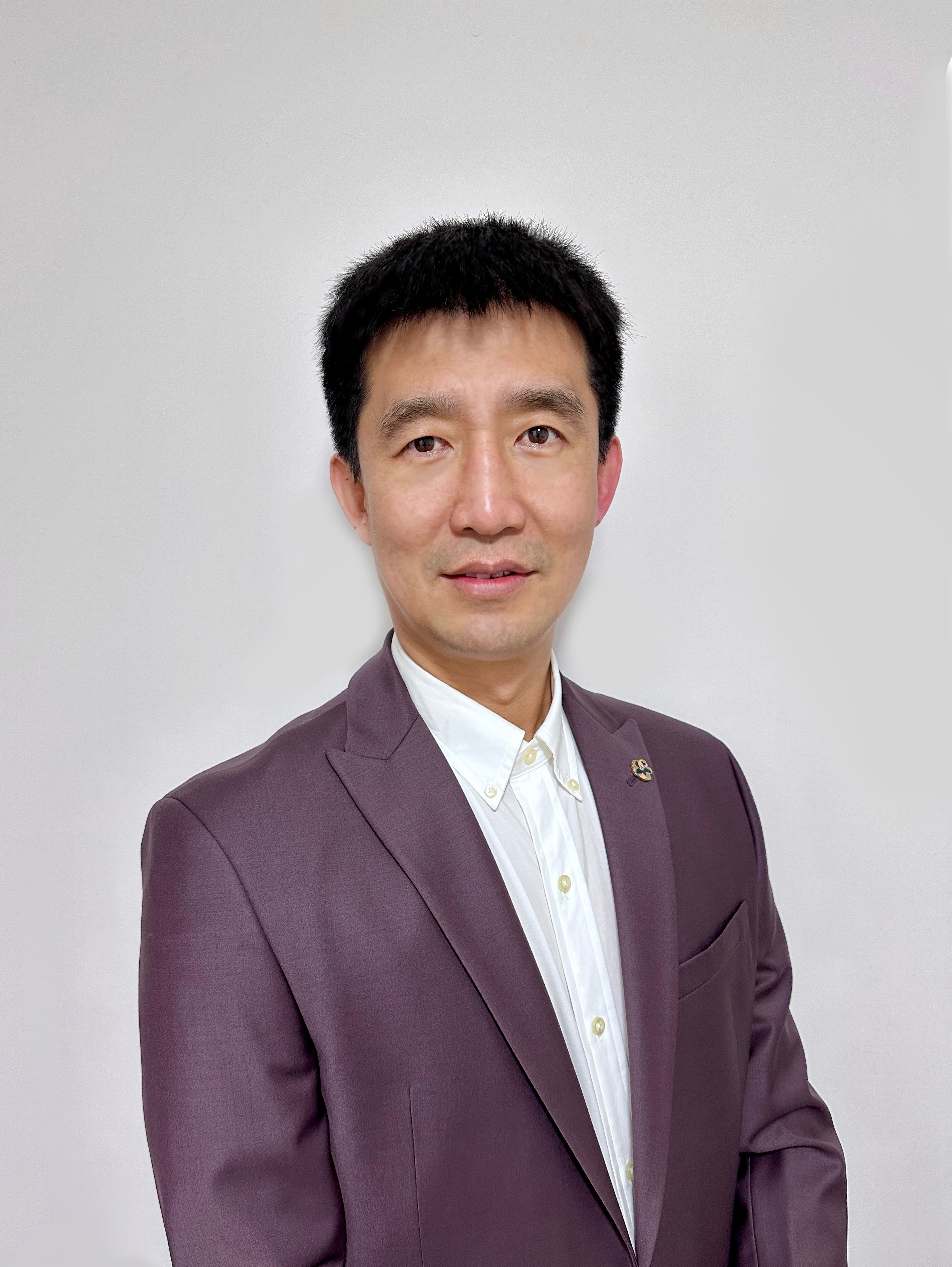}}]{Hongjian Sun} (Senior Member, IEEE) received the Ph.D. degree in electronic and electrical engineering from The University of Edinburgh, U.K., in 2011. He held post-doctoral positions with King’s College London, U.K., and Princeton University, USA. Since 2013, he has been with the University of Durham, U.K., as a Professor (Chair) in smart grid, where he was an Assistant Professor from 2013 to 2017 and an Associate Professor (Reader) from 2017 to 2020. He has authored or coauthored over 180 articles in refereed journals and conferences. He has made contributions to and coauthored the IEEE 1900.6a-2014 Standard. He has authored or coauthored five book chapters and edited two books Smarter Energy: From Smart Metering to the Smart Grid (IET) and From Internet of Things to Smart Cities: Enabling Technologies (CRC). His research mainly focuses on next generation communications and networking, smart grid demand side management and demand response, and renewable energy sources integration. He also served as a Guest Editor for the IEEE Communications Magazine and the IEEE Transactions on Industrial Informatics for several feature topics. He is an Editor-in-Chief for IET Smart Grid journal and an Editor for Journal of Communications and Networks.
\end{IEEEbiography}

\begin{IEEEbiography}[{\includegraphics[width=1in,height=1.25in,clip,keepaspectratio]{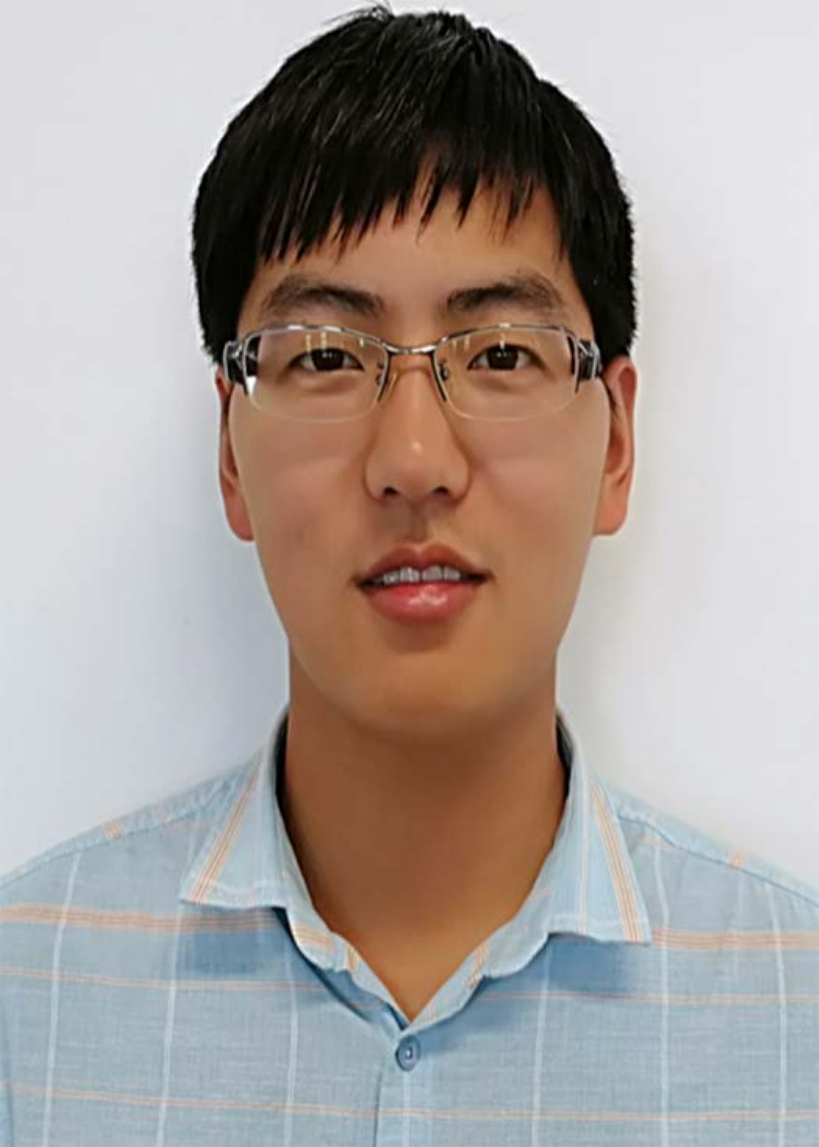}}]{Pengfei Jiao}
received the Ph.D. degree in computer science from Tianjin University, Tianjin, China, in 2018. From 2018 to 2021, he was a lecturer with the Center of Biosafety Research and Strategy of Tianjin University. He is currently a Professor with the School of Cyberspace, Hangzhou Dianzi University, Hangzhou, China.
His current research interests include complex network analysis and its applications.
\end{IEEEbiography}

\end{document}